\documentclass[12pt, draftclsnofoot, journal, onecolumn]{IEEEtran}
\usepackage{blindtext}
\usepackage[T1]{fontenc}
\usepackage{textcomp}
\usepackage{amsmath}
\usepackage{amssymb}
\usepackage{bm}
\usepackage{mdwmath}
\usepackage{cases}
\usepackage{graphicx}
\usepackage{xcolor}
\definecolor{myblue}{rgb}{0,0.4980,1} % Azure
\definecolor{myred}{rgb}{0.8706,0.1608,0.0627} % Chinese red
\usepackage[caption=false,font=footnotesize,subrefformat=parens]{subfig}
\usepackage{makecell}
\usepackage{multirow}
\usepackage{acronym}
\usepackage[nosort,noadjust]{cite}
\usepackage{url}

\usepackage{hyperref}
\hypersetup{%
	bookmarksopen=true,%
	bookmarksnumbered=true,%
	pdfpagemode={UseOutlines},%default
	pdfstartview={FitH},%
	colorlinks=true,%
	linkcolor={myred},%
	citecolor={cyan}}
\usepackage{algpseudocode}
\newcounter{MYalgorithmic}
\renewcommand{\theMYalgorithmic}{\arabic{MYalgorithmic}}
\newcommand{\algcaption}[1]{%
	\refstepcounter{MYalgorithmic}%
	\textbf{Algorithm}~\textbf{\theMYalgorithmic}.~#1}
\newenvironment{MYalgorithmic}[5]
{
	\hrule height 1.2pt
	\vspace{3pt}
	#1{#2}%
	#3{#4}
	\vspace{3pt}
	\hrule height 0.5pt
	\vspace{3pt}
	#5
}
{
	\vspace{3pt}
	\hrule height 0.5pt
}

\newcounter{MYitem}[MYalgorithmic]
\renewcommand{\theMYitem}{\arabic{MYitem}}
\newcommand{\algitem}{%
	\refstepcounter{MYitem}%
	\textbf{\theMYitem)}}

\makeatletter
\newcommand{\MYlabel}[1]{\def\@currentlabel{\theALG@line}\label{#1}}
\makeatother

\newtheorem{problem}{\textbf{Problem}}
\newtheorem{lemma}{\textbf{Lemma}}
\newtheorem{theorem}{\textbf{Theorem}}
\newtheorem{remark}{\textbf{Remark}}
\newtheorem{corollary}{\textbf{Corollary}}
\newtheorem{subproblem}{\textbf{Sub-Problem}}

\newcommand{\upperroman}[1]{\uppercase\expandafter{\romannumeral#1}}

\newcommand{\myvec}[1]{\bm{\mathrm{#1}}}
\newcommand{\myunit}[1]{$ \mathrm{#1} $}
\newcommand{\myexp}{\mathrm{e}}
\DeclareMathOperator{\sbjto}{s.t.}
\DeclareMathOperator{\tr}{tr}
\DeclareMathOperator{\rank}{rank}
\DeclareMathOperator{\diag}{diag}

\newcommand{\MYnewpage}{%
	\ifCLASSOPTIONonecolumn%
		\ifCLASSOPTIONjournal%
			\typeout{The onecolumn journal mode.}%
			\newpage%
		\fi%
	\fi}
\begin{document}
%% *************************************************************************
\ifCLASSOPTIONonecolumn
	\typeout{The onecolumn mode.}
	\title{\LARGE Leveraging Linear Quadratic Regulator Cost and Energy Consumption for Ultra-Reliable and Low-Latency IoT Control Systems}
%	\author{{\normalsize Haojun~Yang,~\MYIEEEmembership{Student Member,~IEEE}}% <-this % stops a space
	\author{Haojun~Yang,~\IEEEmembership{Student Member,~IEEE}, Kuan~Zhang,~\IEEEmembership{Member,~IEEE}, Kan~Zheng,~\IEEEmembership{Senior Member,~IEEE}, and~Yi~Qian,~\IEEEmembership{Fellow,~IEEE}% <-this % stops a space
%		\thanks{Manuscript received XXX, 2019; revised XXX.}% <-this % stops a space
		\thanks{Haojun~Yang and Kan~Zheng are with the Intelligent Computing and Communications ($ \text{IC}^\text{2} $) Lab, Wireless Signal Processing and Networks (WSPN) Lab, Key Laboratory of Universal Wireless Communications, Ministry of Education, Beijing University of Posts and Telecommunications (BUPT), Beijing, 100876, China (E-mail: \textsf{yanghaojun.yhj@bupt.edu.cn; zkan@bupt.edu.cn}).}% <-this % stops a space
		\thanks{Haojun Yang, Kuan~Zhang and Yi~Qian are with the Department of Electrical and Computer Engineering, University of Nebraska-Lincoln (UNL), Omaha, NE 68182, USA  (E-mail: \textsf{haojun.yang@unl.edu; kuan.zhang@unl.edu; yi.qian@unl.edu}).}% <-this % stops a space
%		\thanks{This work was supported by the National Natural Science Foundation of China under Grant No. 61731004.}% <-this % stops a space
	}
\else
	\typeout{The twocolumn mode.}
	\title{Leveraging Linear Quadratic Regulator Cost and Energy Consumption for Ultra-Reliable and Low-Latency IoT Control Systems}
	\author{Haojun~Yang,~\IEEEmembership{Student Member,~IEEE}, Kuan~Zhang,~\IEEEmembership{Member,~IEEE}, Kan~Zheng,~\IEEEmembership{Senior Member,~IEEE}, and~Yi~Qian,~\IEEEmembership{Fellow,~IEEE}% <-this % stops a space
%		\thanks{Manuscript received XXX, 2019; revised XXX.}% <-this % stops a space
		\thanks{Haojun~Yang and Kan~Zheng are with the Intelligent Computing and Communications ($ \text{IC}^\text{2} $) Lab, Wireless Signal Processing and Networks (WSPN) Lab, Key Laboratory of Universal Wireless Communications, Ministry of Education, Beijing University of Posts and Telecommunications (BUPT), Beijing, 100876, China (E-mail: \textsf{yanghaojun.yhj@bupt.edu.cn; zkan@bupt.edu.cn}).}% <-this % stops a space
		\thanks{Haojun Yang, Kuan~Zhang and Yi~Qian are with the Department of Electrical and Computer Engineering, University of Nebraska-Lincoln (UNL), Omaha, NE 68182, USA  (E-mail: \textsf{haojun.yang@unl.edu; kuan.zhang@unl.edu; yi.qian@unl.edu}).}% <-this % stops a space
%		\thanks{This work was supported by the National Natural Science Foundation of China under Grant No. 61731004.}% <-this % stops a space
	}
\fi

\ifCLASSOPTIONonecolumn
	\typeout{The onecolumn mode.}
\else
	\typeout{The twocolumn mode.}
	\markboth{IEEE Internet of Things Journal
	}{Yang \MakeLowercase{\textit{et al.}}: Title}
\fi

\maketitle

\ifCLASSOPTIONonecolumn
	\typeout{The onecolumn mode.}
	\vspace*{-50pt}
\else
	\typeout{The twocolumn mode.}
\fi
\begin{abstract}
To efficiently support the real-time control applications, networked control systems operating with ultra-reliable and low-latency communications (URLLCs) become fundamental technology for future Internet of things (IoT). However, the design of control, sensing and communications is generally isolated at present. In this paper, we propose the joint optimization of control cost and energy consumption for a centralized wireless networked control system. Specifically, with the ``sensing-then-control'' protocol, we first develop an optimization framework which jointly takes control, sensing and communications into account. In this framework, we derive the spectral efficiency, linear quadratic regulator cost and energy consumption. Then, a novel performance metric called the \textit{energy-to-control efficiency} is proposed for the IoT control system. In addition, we optimize the energy-to-control efficiency while guaranteeing the requirements of URLLCs, thereupon a general and complex max-min joint optimization problem is formulated for the IoT control system. To optimally solve the formulated problem by reasonable complexity, we propose two radio resource allocation algorithms. Finally, simulation results show that our proposed algorithms can significantly improve the energy-to-control efficiency for the IoT control system with URLLCs.
\end{abstract}

\ifCLASSOPTIONonecolumn
	\typeout{The onecolumn mode.}
	\vspace*{-10pt}
\else
	\typeout{The twocolumn mode.}
\fi
\begin{IEEEkeywords}
Internet of things (IoT), networked control systems, ultra-reliable and low-latency communications (URLLC), energy-to-control efficiency (ECE), finite blocklength theory, massive MIMO.
\end{IEEEkeywords}

\IEEEpeerreviewmaketitle

\MYnewpage

%\renewcommand{\IEEEiedlistdecl}{\IEEEsetlabelwidth{DDDDDDDD}}
%\begin{acronym}
%	\acro{ITS} {Intelligent Transportation System}
%	\acro{LTE} {Long Term Evolution}
%\end{acronym}
%\renewcommand{\IEEEiedlistdecl}{\relax} % remember to reset \IEEEiedlistdecl
%% *************************************************************************

\section{Introduction}
\label{sec:Introduction}

\IEEEPARstart{R}{ecent} advances in computing, communications, control and sensing promote the rapid development of Internet of things (IoT)~\cite{Fuqaha2015,Chiang2016}. Some emerging control applications of IoT, such as autonomous driving, ``Industry 4.0'' and tactile Internet, are gradually changing people's lives~\cite{Kuutti2018,Antonakoglou2018,Aceto2019}. The evolution and realization of these emerging applications heavily rely on the support of networked control systems~\cite{Park2018}. In general, a networked control system is established by four basic elements, namely sensors, controllers, actuators and communication networks. Its control loops are closed through a communication network. With the aid of networked control systems, the real-time control can be provided for physical plants. For example, except sophisticated sensors, autonomous driving is more in need of robust communication networks exchanging real-time information to offer the assistance of control~\cite{Kuutti2018}. Therefore, networked control systems become fundamental technology for the future IoT.

The design of networked control systems should leverage three vital aspects, i.e., latency, reliability and cost. First of all, timely and precise control commands are usually required by some of IoT control applications. For instance, in order to control various plants in ``Industry 4.0'', factory automation may requires a transmission latency within a few milliseconds (1$ \sim $5~\myunit{ms}) and a reliability in terms of error probability down to $ 10^{-5} $ (or $ 10^{-6} $)~\cite{Aceto2019}. To meet these rigorous requirements, ultra-reliable and low-latency communications (URLLCs) become an indispensable component for networked control systems~\cite{itu2015}. In addition, reducing cost is also important for IoT control systems. System costs consist of control cost and energy consumption. In the control theory, the control cost is also referred to as the ``$ J $-function'' or the linear quadratic regulator (LQR) cost~\cite{Dorfbook}. For example, an important control goal is to minimize the mean square deviation between the plant state and the desired state $ \myvec{0} $. As a result, minimizing the LQR cost is equivalent to stabilizing plants~\cite{Tatikonda2004,Nair2007}. Meanwhile, the lower energy consumption is more beneficial to let plants run on small and inexpensive batteries for up to many years, which can achieve the goal of green communications~\cite{Xiong2015}. Hence, it is crucial to balance the tradeoff among latency, reliability and cost for IoT control systems.

Some recent works study the tradeoff between latency and reliability from the aspects of the physical layer and media access control (MAC) layer~\cite{YangMag2019,Mei2018TWC}. Compared with the queueing latency of the MAC layer, the transmission latency of the physical layer may be more crucial for URLLCs. The transmission latency mainly depends on the frame design of the physical layer~\cite{YangMag2019,3gpp913}. With respect to the other works in cost, stabilizing plants is equivalent to minimizing the control cost in the case of mean square deviation~\cite{Tatikonda2004,Nair2007}. Thereupon, the optimization of the control cost is investigated for the vehicular platooning and industrial IoT~\cite{Mei2018TVT,Lyu2018}. Moreover, the higher transmission power is supported, the better performance of latency, reliability and control is achieved, but it comes at the cost of higher energy consumption. To address this issue, the joint optimization of energy consumption and control system performance is investigated in~\cite{Sadi2017}, where the control performance is not modeled as the common LQR cost.

However, the existing works still have the some limitations for the design of the networked control systems with URLLCs. First of all, it is essential to take the requirements of URLLCs into consideration from the aspect of the physical layer. This is because the other types of latency can be significantly reduced by exploiting the customized network architecture, such as the network slicing. In addition, the joint optimization of control, sensing and communications may be more in line with actual needs for the future IoT control systems dedicated for the emerging real-time applications. The joint optimization is also conducive to making the IoT control system design more universal. To this end, it is paramount to jointly optimize the LQR cost and energy consumption for the networked control systems operating with URLLCs.

In this paper, we investigate the joint optimization of the LQR cost and energy consumption for the IoT control systems operating with URLLCs. Specifically, considering a centralized wireless networked control system, we develop a optimization framework of the LQR cost and the round-trip energy consumption. Based on the proposed framework, we optimize the performance of energy-to-control efficiency while guaranteeing the requirements of URLLCs. The main contributions of this paper are summarized as follows.
\begin{itemize}
\item Considering a centralized wireless networked control system with URLLCs, we sort out the relationship among control, sensing and communications in detail.

\item With the ``sensing-then-control'' protocol, we develop a optimization framework for the networked control system with URLLCs. In this framework, we first derive the spectral efficiency, LQR cost and energy consumption. Then, we propose a novel performance metric called the energy-to-control efficiency. The rationality and validity of the proposed performance metric are also proved.

\item Based on the proposed framework, we formulate a general and complex max-min joint optimization problem. Two radio resource allocation algorithms are put forward to optimally solved the formulated problem by reasonable complexity.

\item Our simulation results show that the proposed algorithms can significantly improve the energy-to-control efficiency, while guaranteeing the requirements of URLLCs for the networked control system.
\end{itemize}

The remainder of this paper is organized as follows. First of all, Section~\ref{sec:Related} reviews the related works of networked control systems, and Section~\ref{sec:Model} describes the centralized wireless networked control system with URLLCs. Then, Section~\ref{sec:Framework} proposes the optimization framework, and the joint optimization of the LQR cost and energy consumption is studied in Section~\ref{sec:Optimization}. Finally, Section~\ref{sec:Simulation} illustrates the simulation results, while the conclusions are offered in Section~\ref{sec:Conclusion}.

\textit{Notations}: Uppercase boldface letters and lowercase boldface letters denote matrices and vectors, respectively, while $ \myvec{I}_N $ denotes an $ N \times N $ identity matrix. $ (\cdot)^\text{T} $, $ (\cdot)^\text{*} $, $ (\cdot)^\text{H} $ and $ (\cdot)^{-1} $ represent the transpose, conjugate, conjugate transpose and pseudo-inverse of a matrix/vector, respectively. $ \diag\lbrace \myvec{a} \rbrace $ is a diagonal square matrix whose main diagonal is formed by the vector $ \myvec{a} $, while $ \det(\cdot) $ and $ \tr(\cdot) $ denote the determinant and trace of a square matrix, respectively. $ \rank(\cdot) $ denotes the rank of a matrix. Moreover, $ \nabla $ represents the gradient. Finally, $ \mathbb{E}(\cdot) $ represents the mathematical expectation, while $ \mathbb{CN}(\mu,\sigma^2) $ is the complex Gaussian distribution with mean $ \mu $ and real/imaginary component variance $ \sigma^2/2 $.

\section{Related Works}
\label{sec:Related}

The existing works about networked control systems mainly focus on the aspects of state sensing, plant stability, control cost, communication latency and reliability, as well as energy consumption~\cite{Park2018,Zhang2016,Nair2007}.

\subsubsection{State Sensing}
In general, the continuous-time plant states need to be sensed (or sampled) before transmitting them over wireless communication networks. There are two methods to sense continuous-time states, namely the time-triggered method and event-triggered method~\cite{Heemels2012}. In the time-triggered method, sensors sample the plant states based on the pre-specified sensing duration and period. In the event-triggered method, sensing actions can only be performed when the stability or pre-specified control performance are about to lost. Hence, compared with the time-triggered method, the event-triggered method can reduce the system overhead~\cite{Ara2014}. However, by exploiting the efficient resource allocation for wireless communication networks, the time-triggered sensing is more beneficial to design the networked control systems with URLLCs~\cite{Ergen2006}. Based on the idea of time-triggered sensing, the optimization of sensing duration is studied for wireless control systems with massive MIMO~\cite{Zhao2017}. In addition, with the goal of minimizing the average distortion, the problem of sensing accuracy is investigated from the perspective of the rate distortion theory~\cite{Orhan2015}. Thus, this paper also adopts the idea of time-triggered sensing to meet the requirements of URLLCs.

\subsubsection{Plant Stability and Control Cost}
The stability of plants is one of the fundamental requirements for networked control systems. Generally, the stability of plants is strongly related to the control cost. The control cost can be quantified as a function of plant states and control inputs~\cite{Dorfbook}. As previously mentioned, for the case of the mean square deviation, stabilizing plants is equivalent to minimizing the control cost~\cite{Tatikonda2004,Nair2007}. The tradeoff study between the control cost and communication constraints can be traced in~\cite{Tatikonda2004}, where the necessary condition for stabilizing a vector linear plant with bounded noise is illustrated. Following the above study, the minimum information capacity required to maintain the plant at a pre-specified LQR cost is revealed in~\cite{Kostina2019}. For the practical systems, the optimization of the control cost is investigated for the vehicular platooning and industrial IoT~\cite{Mei2018TVT,Lyu2018}.

\subsubsection{Communication Latency and Reliability}
According to the hierarchical architecture of networks, the communication latency is generally divided into the transmission latency of the physical layer and the queueing latency of the MAC layer~\cite{YangMag2019}. The transmission latency mainly depends on the frame design of the physical layer~\cite{YangMag2019,3gpp913}. The problems of minimizing the queueing latency are studied in~\cite{Sun2016,Mei2018TWC,Zheng2016,Zheng2015}, where the reliability is modeled as the outage probability or the block error rate. Compared with the queueing latency, the transmission latency may be more crucial for URLLCs. This is because the queueing latency can be significantly reduced by exploiting the customized network architecture, such as the network slicing~\cite{Ksentini2018}. For the transmission latency, both the ergodic capacity and the outage capacity are no longer applicable, since they violate the requirements of URLLCs~\cite{Giuseppe2016}. Thus, the finite blocklength theory becomes a powerful technique to address the URLLC-related optimization problems~\cite{Polyanskiy2010,She2017}.

\subsubsection{Energy Consumption}
Energy consumption is another important aspect for networked control systems. Increasing the transmission power can improve the signal-to-interference-plus-noise ratio (SINR), thereby reducing the transmission latency and enhancing the reliability~\cite{YangMag2019,Mei2018TWC,zhengsurvey2015MIMO,Liu2014}. However, this comes at the cost of higher energy consumption. In order to deal with this issue, the joint optimization of energy consumption and control system performance is investigated in~\cite{Sadi2017}, where the control performance is not modeled as the common LQR cost.

To sum up, the existing works may have the following limitations for the design of networked control systems with URLLCs, i.e.,
\begin{itemize}
\item The existing works rarely take the requirements of URLLCs into consideration from the perspective of the physical layer, which is essential for the networked control systems dedicated for the emerging real-time IoT applications.

\item At present, the design of control, sensing and communications is isolated. The existing works mainly focus on one of the above topics. The joint optimization of the above four aspects may be more in line with actual needs for the future IoT control systems.
\end{itemize}

In summary, the joint optimization of control, sensing and communications is an open research problem for the networked control systems with URLLCs.

\section{Centralized Wireless Networked Control System Model}
\label{sec:Model}

\begin{figure}[!t]
	\centering
	\includegraphics[scale=0.5]{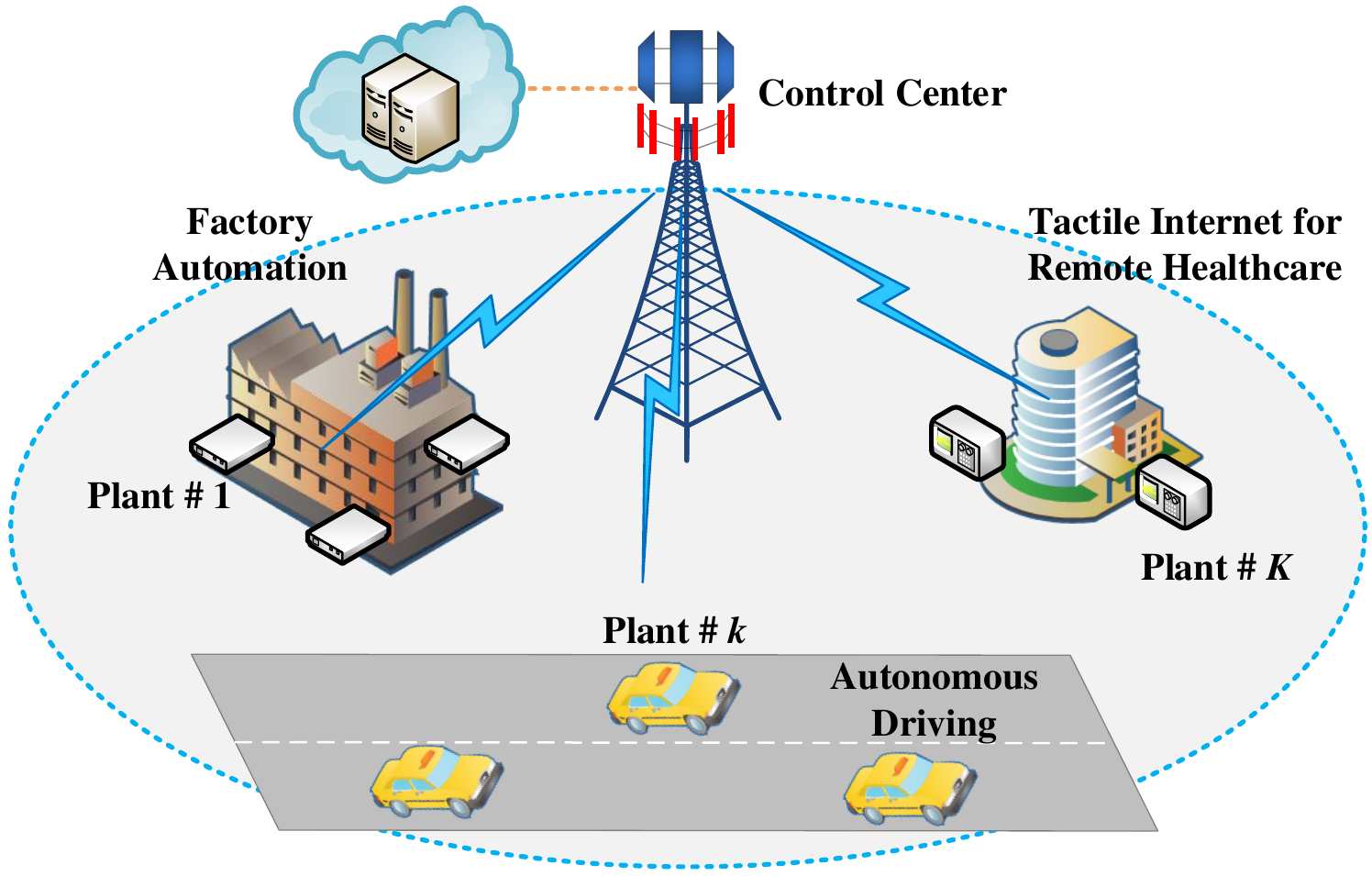}
	\caption{Centralized wireless networked control system model.}
	\label{fig_Model}
\end{figure}

\begin{figure}[!t]
	\centering
	\includegraphics[scale=0.6]{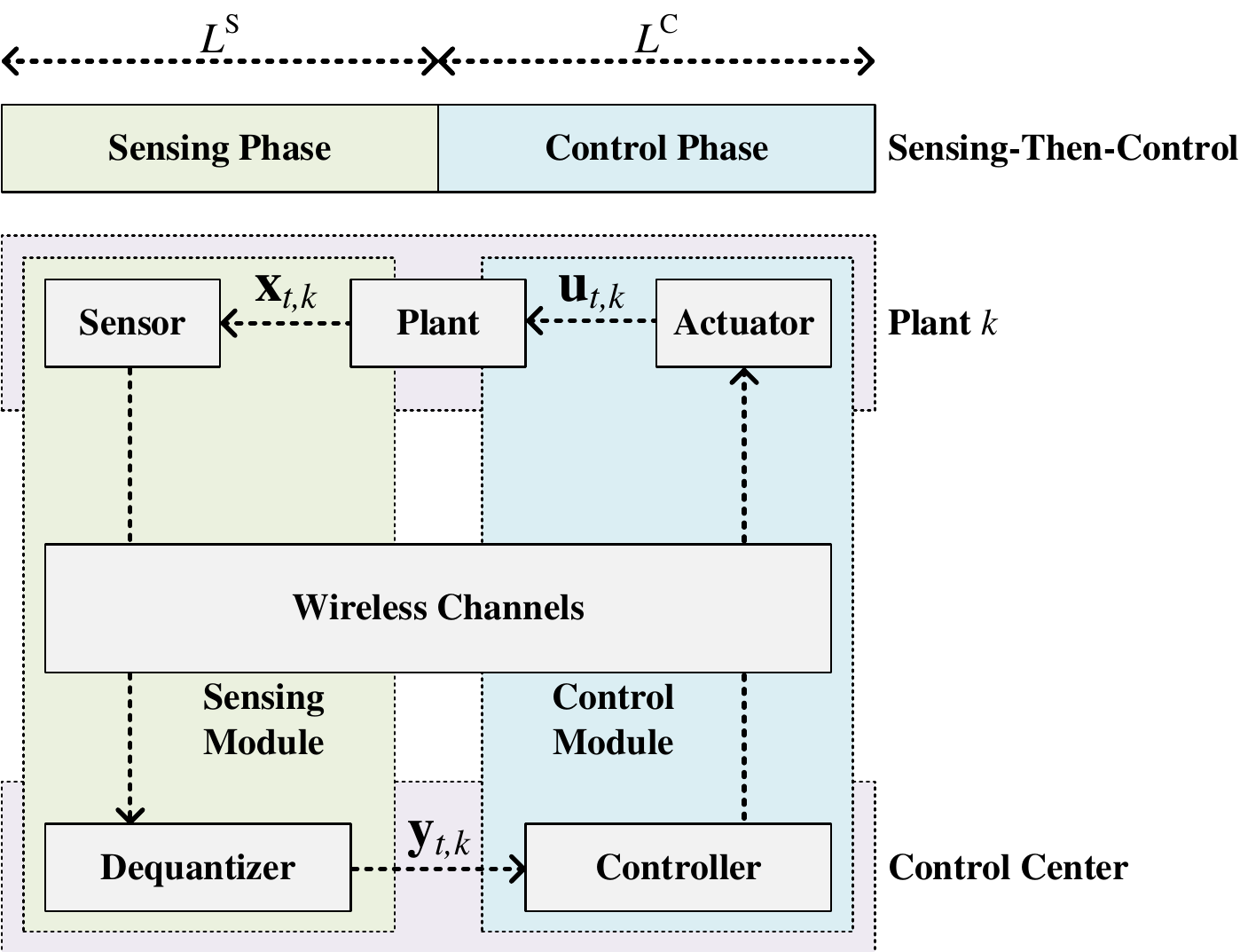}
	\caption{Sensing phase and control phase for a plant in centralized wireless networked control system.}
	\label{fig_Module}
\end{figure}

The detailed centralized wireless networked control system model is established in this section.

\subsection{System Description}
As shown in Fig.~\ref{fig_Model}, we consider a centralized wireless networked control system with URLLCs in this paper. The single-cell system consists of a control center and $ K $ plants. Each plant has a actuator and a sensor, while the control center includes a controller and a dequantizer~\cite{Mahmoud2018,Zhang2016}. The control center, which can be also viewed as a base station, employs $ M $ antennas and simultaneously communicates with $ K~(M \gg K) $ single-antenna plants. Moreover, in order to reduce the overhead of channel estimation, the system operates at the time-division duplex (TDD) mode. Finally, the ``\textit{sensing-then-control}'' protocol is performed for the system. More specifically, the whole procedure is divided into two phases. In Phase 1, each sensor reports its sensing information to the control center. While in Phase 2, the control center transmits the corresponding control actions to all actuators based on the sensing information.

\subsection{Channel Model}
Since the system operates at the TDD mode, the channel reciprocity holds for all links during two phases. All channels experience independent flat block-fading, i.e., they remain constant during a coherence block (time-bandwidth product), but change independently from one block to another. Let $ \myvec{g}_k = \beta_k^{1/2} \myvec{h}_k \in \mathbb{C}^{M \times 1} $ be the channel vector between the $ k $-th plant and the control center, where $ \beta_k $ and $ \myvec{h}_k $ denote the large-scale fading and small-scale fading. In addition, in this paper, imperfect channel estimation is considered~\cite{Kashyap2016,Zhao2017}, whereupon we have
\begin{align}
\myvec{g}_k = \beta_k^{1/2}\myvec{h}_k = \sqrt{\beta_k \chi_k}\hat{\myvec{h}}_k + \sqrt{\beta_k \left( 1-\chi_k \right)}\myvec{e}_k, \forall k,
\end{align}
where $ \hat{\myvec{h}}_k $, $ \myvec{e}_k $ and $ \chi_k \in [0,1] $ are the estimate, error and estimation accuracy of $ \myvec{h}_k $, respectively. Finally, each element of $ \hat{\myvec{h}}_k $ and $ \myvec{e}_k $ is independent and identically distributed (i.i.d.) complex Gaussian random variable with mean 0 and variance 1, namely $ \hat{\myvec{h}}_k, \myvec{e}_k \mathop \sim \mathbb{CN}(\myvec{0},\myvec{I}_M) $.

\subsection{Sensing Phase}
As shown in Fig.~\ref{fig_Module}, in the sensing phase, each sensor acquires the real-time information on its own plant state and sends them to the control center. 

\subsubsection{Sensing Module}
The goals of the sensing module are to acquire all real-time information on the plant state and send them to the control center. However, the condition of $ \myvec{y}_{t,k} = \myvec{x}_{t,k} $ is too rigorous for practical engineering implementations. To this end, in this paper, the plant is defined as ``fully observed'', when the condition $ \Pr\left[ D(\myvec{x}_{t,k}, \myvec{y}_{t,k}) > d_k \right] \leqslant \delta_k $ holds for the given small values of $ d_k $ and $ \delta_k $. $ D(\myvec{x}_{t,k}, \myvec{y}_{t,k}) $ denotes the distortion function here, and $ d_k $ is the distortion level. A common solution to determine the minimum information capacity of describing the plant state is the classical rate-distortion theory, whereupon the rate-distortion function $ R^\text{S}_k(d_k) $ can be used between the sensor and the dequantizer. In addition, given the channel capacity from the sensor to the dequantizer $ C^\text{S2D}_k $, according to the source-channel separation theorem with distortion~\cite{Coverbook}, we have $ C^\text{S2D}_k > R^\text{S}_k(d_k) $ for the design of the sensing module. Nevertheless, both $ C^\text{S2D}_k $ and $ R^\text{S}_k(d_k) $ can only be approached at the cost of excessive latency (blocklengths) and complexity. Therefore, in order to meet the requirements of URLLCs, based on the finite blocklength theory, the tradeoff between $ C^\text{S2D}_k $ and $ R^\text{S}_k(d_k) $ can be well approximated by~\cite{Kostina2013}:
\begin{align}
\label{setradoff}
L^\text{S} B C^\text{S2D}_k - N_k R^\text{S}_k\left( d_k \right) \approx \sqrt{L^\text{S} B V^\text{S}_k + N_k W^\text{S}_k} Q^{-1}\left( \delta_k \right), \forall k,
\end{align}
where $ L^\text{S} $ is the \textit{transmission latency} for the sensing phase, and $ B $ is the system bandwidth. $ L^\text{S} B $, which is also referred to as the number of transmission symbols, represents the channel uses. $ N_k $ is the number of the plant states. $ \delta_k $ is the proxy for the \textit{sensing-transmission reliability} (generally $ 10^{-5} $ or $ 10^{-6} $), and $Q^{-1}(\cdot) $ denotes the inverse of the Gaussian $ Q $-function. $ V^\text{S}_k $ and $ W^\text{S}_k $ are the so-called channel dispersion and rate-dispersion, respectively. According to~\eqref{setradoff}, one can conclude that $ C^\text{S2D}_k > R^\text{S}_k(d_k) $ still holds when both $ L^\text{S} B $ and $ N_k $ tend to infinity. Consequently,~\eqref{setradoff} is more beneficial to optimize latency and reliability for URLLC systems.

\subsubsection{Communication Model}
The received signal at the the control center $ \myvec{y}^\text{S} \in \mathbb{C}^{M \times 1} $ can be written as
\begin{align}
\myvec{y}^\text{S} = \sum_{k=1}^{K} \sqrt{\dfrac{p^\text{S}_{k}}{M}}\myvec{g}_{k}s^\text{S}_{k} + \myvec{n}^\text{S},
\end{align}
where $ p^\text{S}_{k} $ is the transmission power of the $ k $-th plant during the sensing phase, and $ \myvec{n}^\text{S} $ is the i.i.d. complex additive white Gaussian noise (AWGN) with $ \myvec{n}^\text{S} \sim \mathbb{CN}(\myvec{0},\sigma^2_\text{S} \myvec{I}_M) $. $ s^\text{S}_{k} \in \mathbb{C} $ denotes the date symbol of the $ k $-th plant, where $ \mathbb{E}\left[ |s^\text{S}_{k}|^2 \right] = 1 $. It is widely exploited that low-complexity linear detection techniques are capable of asymptotically attaining optimal performance in massive MIMO~\cite{zhengsurvey2015MIMO}. Therefore, we adopt the low-complexity maximum ratio combining (MRC) detection in this paper. Recall from $ \hat{\myvec{g}}_k = \beta_k^{1/2}\hat{\myvec{h}}_k $ then the post-processing signal of $ \myvec{y}^\text{S} $ can be expressed as
\begin{align}
\hat{s}^\text{S}_{k} &= \hat{\myvec{g}}_{k}^\text{H} \myvec{y}^\text{S} \notag \\
&= \sqrt{\dfrac{p^\text{S}_{k}}{M}}\sqrt{\chi_k} \hat{\myvec{g}}_{k}^\text{H} \hat{\myvec{g}}_k s^\text{S}_{k} + \sum_{\substack{i=1, \\ i \neq k}}^{K} \sqrt{\dfrac{p^\text{S}_{i}}{M}}\sqrt{\chi_i} \hat{\myvec{g}}_{k}^\text{H} \hat{\myvec{g}}_i s^\text{S}_{i} \notag \\
&\quad + \sum_{j=1}^{K} \sqrt{\dfrac{p^\text{S}_{j}}{M}}\sqrt{1-\chi_j}\beta_j^{1/2}\hat{\myvec{g}}_{k}^\text{H} \myvec{e}_j s_j^\text{S} + \hat{\myvec{g}}_{k}^\text{H} \myvec{n}^\text{S}.
\end{align}
As a result, the SINR of the $ k $-th plant during the sensing phase is given by
\begin{align}
\gamma_k^\text{S} = \dfrac{\dfrac{p^\text{S}_{k}}{M}\chi_k \left\| \hat{\myvec{g}}_k \right\|^4}{\sum\limits_{\substack{i=1, \\ i \neq k}}^{K} \dfrac{p^\text{S}_{i}}{M}\chi_i \left| \hat{\myvec{g}}_{k}^\text{H} \hat{\myvec{g}}_i \right|^2 + \left\| \hat{\myvec{g}}_k \right\|^2 \sum\limits_{j=1}^{K} \dfrac{p^\text{S}_{j}}{M} \beta_j \left( 1-\chi_j \right) + \sigma^2_\text{S}\left\| \hat{\myvec{g}}_k \right\|^2}.
\end{align}

\subsection{Control Phase}
As shown in Fig.~\ref{fig_Module}, in the control phase, various control actions are modulated into the transmission symbols and sent to each actuator, then each actuator performs the control actions sent by the controller, in order to stabilize its own plant.

\subsubsection{Control Module}
In the control theory, a discrete-time linear stochastic dynamical plant is generally given by
\begin{align}
\label{plant}
\myvec{x}_{t+1,k} = \myvec{A}_k \myvec{x}_{t,k}+\myvec{B}_k \myvec{u}_{t,k}+\myvec{w}_{t,k}, \forall t \geqslant 0, k,
\end{align}
where $ \myvec{x}_{t+1,k} \in \mathbb{R}^{N_k \times 1} $ and $ \myvec{x}_{t,k} \in \mathbb{R}^{N_k \times 1} $ denote the plant state vector at time $ t+1 $ and $ t $, respectively. $ \myvec{u}_{t,k} \in \mathbb{R}^{N_k \times 1} $ represents the control input for the actuator, and $ \myvec{w}_{t,k} \in \mathbb{R}^{N_k \times 1} $ represents the noise for the control process. $ \myvec{A}_k \in \mathbb{R}^{N_k \times N_k} $ and $ \myvec{B}_k \in \mathbb{R}^{N_k \times N_k} $ are the fixed matrices, and they are a controllable pair~\cite{Dorfbook,Nair2007}. As illustrated in~\cite{Nair2007}, the stable eigenvalues of $ \myvec{A}_k $ have no effects on control. Therefore, we let $ \myvec{A}_k $ only have the unstable eigenvalues in this paper. Furthermore, $ \myvec{A}_k $ and $ \myvec{B}_k $ are full rank matrices, namely $ \rank(\myvec{A}_k) = \rank(\myvec{B}_k) = N_k $.

A common performance metric on evaluating control efficiency at time $ T $ is given by
\begin{align}
J_{T,k} = \mathbb{E}\left[ \sum_{t=1}^{T-1} \left( \myvec{x}_{t,k}^\text{T} \myvec{Q}_k \myvec{x}_{t,k} + \myvec{u}_{t,k}^\text{T} \myvec{R}_k \myvec{u}_{t,k} \right) + \myvec{x}_{T,k}^\text{T} \myvec{S}_{T,k} \myvec{x}_{T,k} \right],
\end{align}
where $ \myvec{Q}_k $, $ \myvec{R}_k $ and $ \myvec{S}_{T,k} $ are the positive semi-definite matrices. $ J_{T,k} $ is also referred to as the \textit{\textbf{LQR cost function}}. As an important and special case, let $ \myvec{Q}_k = \myvec{S}_{T,k} = \myvec{I}_{N_k} $ and $ \myvec{R}_k = \myvec{0} $, then $ J_{T,k} = \mathbb{E}\left[ \sum_{t=1}^{T} || \myvec{x}_{t,k} ||^2 \right] $ which denotes the mean square deviation between the plant state $ \myvec{x}_{t,k} $ and the desired state $ \myvec{0} $. For the mean square deviation, the definition of plant stability is to make $ \limsup_{t \rightarrow \infty} \mathbb{E}\left[ || \myvec{x}_{t,k} ||^2 \right] $ bounded, namely $ \limsup_{t \rightarrow \infty} \mathbb{E}\left[ || \myvec{x}_{t,k} ||^2 \right] < \infty $~\cite{Tatikonda2004}. Furthermore, in order to model the fundamental limits of control under communication constraints more precisely, a rate-cost function $ R^\text{C}_k(c_k) $ is developed in~\cite{Kostina2019}, where the minimum information capacity required to maintain the plant at LQR cost $ c_k $ is revealed for both noiseless and noisy channels. The lemma about the lower bound of the rate-cost function is illustrated as follows.

\begin{lemma}[\cite{Kostina2019}]
\label{lemma1}
For the fully observed dynamical plant \eqref{plant}, $ c_k > c^\text{min}_k $ and positive definite matrix $ \myvec{M}_k $, the lower bound of the rate-cost function is given by
\begin{align}
R^\text{C}_k\left( c_k \right) &\geqslant \log_2 \left| \det \left( \myvec{A}_k \right) \right| \notag \\
&\qquad + \dfrac{N_k}{2}\log_2 \left( 1+ \dfrac{Z\left( \myvec{w}_k \right) \left| \det \left( \myvec{M}_k \right) \right|^\frac{1}{N_k}}{\frac{1}{N_k}\left( c_k - c^\text{min}_k \right)} \right),
\end{align}
where $ c^\text{min}_k = \tr(\myvec{\Sigma}_{\myvec{w},k} \myvec{S}_k) $, $ \myvec{\Sigma}_{\myvec{w},k} $ is the covariance matrix of $ \myvec{w}_{t,k} $, and $ \myvec{S}_k $ is the solution to the algebraic Riccati equations:
\begin{align}
\begin{cases}
\myvec{S}_k = \myvec{Q}_k + \myvec{A}^\text{T}_k\left( \myvec{S}_k - \myvec{M}_k \right) \myvec{A}_k, \\
\myvec{M}_k = \myvec{S}_k \myvec{B}_k\left( \myvec{R}_k + \myvec{B}^\text{T}_k \myvec{S}_k \myvec{B}_k \right)^{-1} \myvec{B}^\text{T}_k \myvec{S}_k.
\end{cases}
\end{align}
$ Z(\cdot) $ denotes the entropy power function, i.e.,
\begin{align}
Z\left( \myvec{w}_k \right) = \dfrac{1}{2\pi\myexp} \exp\left[ \dfrac{2}{N_k} h\left( \myvec{w}_k \right) \right],
\end{align}
and $ h(\myvec{w}_k) > -\infty $ is the the differential entropy of $ \myvec{w}_{t,k} $.
\end{lemma}

Obviously, based on Lemma~\ref{lemma1}, one can find that $ R^\text{C}_k\left( c_k \right) \geqslant \log_2 \left| \det \left( \myvec{A}_k \right) \right| $. Moreover, it is proved that the separated design of control and communications is optimal when the previous control inputs are adopted~\cite{Kostina2019}. Hence, compared with distributed systems, centralized systems are more beneficial to reduce system costs. This is because the centralized systems do not require the feedback of the previous control inputs which can be store at the local memory of the control center.

\subsubsection{Communication Model}
Let $ \myvec{s}^\text{C} = [ s_1^\text{C}, \cdots, s_k^\text{C}, \cdots, s_K^\text{C} ]^\text{T} \in \mathbb{C}^{K \times 1} $ represent the symbol vector, where $ \mathbb{E}\left[ \myvec{s}^\text{C} (\myvec{s}^\text{C})^\text{H} \right]=\myvec{I}_K $. Similarly with the MRC detection, the low-complexity linear matched filter (MF) precoder is adopted in this paper. Denote $ \myvec{v}_k \in \mathbb{C}^{M \times 1} $ as the precoder vector, i.e.,
\begin{align}
\myvec{v}_k = \dfrac{\hat{\myvec{g}}_k^{*}}{\left\| \hat{\myvec{g}}_k^{*} \right\|}.
\end{align}
The received signal of the $ k $-th plant can be written as
\begin{align}
y_k^\text{C} &= \sum_{i=1}^{K}\sqrt{\dfrac{p_i^\text{C}}{M}}\myvec{g}_k^\text{T}\myvec{v}_i s_i^\text{C} + n_k^\text{C} \notag \\
&= \sqrt{\dfrac{p_k^\text{C}}{M}}\sqrt{\chi_k}\hat{\myvec{g}}_k^\text{T}\myvec{v}_k s_k^\text{C}+\sum_{\substack{i=1, \\ i \neq k}}^{K} \sqrt{\dfrac{p_i^\text{C}}{M}}\sqrt{\chi_k}\hat{\myvec{g}}_k^\text{T}\myvec{v}_i s_i^\text{C} \notag \\
&\quad + \sum_{j=1}^{K} \sqrt{\dfrac{p_j^\text{C}}{M}}\sqrt{1-\chi_k}\beta_k^{1/2}\myvec{e}_k^\text{T}\myvec{v}_j s_j^\text{C} + n_k^\text{C},
\end{align}
where $ p_k^\text{C} $ is the transmission power of the $ k $-th plant during the control phase, and $ n_{k}^\text{C} $ is the i.i.d. complex AWGN with $ n_{k}^\text{C} \mathop \sim \mathbb{CN}(0,\sigma^2_\text{C}) $. Therefore, the SINR of the $ k $-th plant during the control phase is given by
\begin{align}
\gamma_k^\text{C} = \dfrac{\dfrac{p_k^\text{C}}{M}\chi_k\left| \dfrac{\hat{\myvec{g}}_k^\text{T}\hat{\myvec{g}}_k^{*}}{\left\| \hat{\myvec{g}}_k^{*} \right\|} \right|^2}{\chi_k \sum\limits_{\substack{i=1, \\ i \neq k}}^{K} \dfrac{p_i^\text{C}}{M} \left| \dfrac{\hat{\myvec{g}}_k^\text{T}\hat{\myvec{g}}_i^{*}}{\left\| \hat{\myvec{g}}_i^{*} \right\|} \right|^2+ \beta_k\left( 1-\chi_k \right) \sum\limits_{j=1}^{K} \dfrac{p_j^\text{C}}{M} + \sigma^2_\text{C}}.
\end{align}
%where the interference term $ I_k^\text{C} $ can be expressed as
%\begin{align}
%I_k^\text{C} = \sum_{\substack{i=1, \\ i \neq k}}^{K} \dfrac{p_i^\text{C}}{M}\chi_k\left| \dfrac{\hat{\myvec{g}}_k^\text{T}\hat{\myvec{g}}_i^{*}}{\left\| \hat{\myvec{g}}_i^{*} \right\|} \right|^2+ \sum_{j=1}^{K} \dfrac{p_j^\text{C}}{M}\beta_k\left( 1-\chi_k \right).
%\end{align}

\section{Optimization Framework of LQR Cost and Energy Consumption}
\label{sec:Framework}

For the networked control system with URLLCs, our optimization objective is to minimize the LQR cost of the control phase, while minimizing the round-trip energy consumption. In this section, we develop a optimization framework of the LQR cost and the round-trip energy consumption. Specifically, we first derive the spectral efficiency (SE), the LQR cost and the energy consumption. Then, we propose a novel performance metric for the networked control system with URLLCs. Furthermore, we prove the rationality and validity of the proposed performance metric.

\subsection{Spectral Efficiency of Sensing Phase and Control Phase}
The theorem about the SE of the sensing phase and the control phase is illustrated as follows.

\begin{theorem}
\label{theo1}
For the large number of antennas $ M $ at the control center, the ergodic SE during two phases can be well approximated by
\begin{align}
\label{SEC}
C^\text{S2D}_k &= \log_2\left( 1+\dfrac{p^\text{S}_{k}\chi_k\beta_k}{\sigma^2_\text{S}} \right), \forall k, \\
\label{SES}
R^\text{C2A}_k &= \log_2\left( 1+\dfrac{p_k^\text{C}\chi_k\beta_k}{\sigma^2_\text{C}} \right) - \sqrt{\dfrac{1}{L^\text{C}B}}Q^{-1}\left( \epsilon_k \right) \log_2 \myexp, \forall k,
\end{align}
where $ L^\text{C} $ is the \textit{transmission latency} for the control phase, and $ \epsilon_k $ is the proxy for the \textit{transmission reliability} during the control phase (generally $ 10^{-5} $ or $ 10^{-6} $).
\end{theorem}

\begin{IEEEproof}
See Appendix~\ref{app:theo1}.
\end{IEEEproof}

Furthermore, we have the following corollary for the sensing phase, i.e.,

\begin{corollary}
For the networked control system with URLLCs, the ergodic SE during the sensing phase must satisfy
\begin{align}
& L^\text{S}B\log_2\left( 1+\dfrac{p^\text{S}_{k}\chi_k\beta_k}{\sigma^2_\text{S}} \right) \geqslant \dfrac{N_k}{2}\log_2\left( \dfrac{\sigma^2_\text{PS}}{d_k} \right) \notag \\
&\qquad \qquad \qquad \quad + \sqrt{L^\text{S}B + \dfrac{N_k}{2}} Q^{-1}\left( \delta_k \right) \log_2 \myexp, \forall k,
\end{align}
where $ \sigma^2_\text{PS} $ is the variance of the plant state, $ d_k $ is the distortion level, and $ \delta_k $ is the proxy for the \textit{sensing-transmission reliability} during the sensing phase (generally $ 10^{-5} $ or $ 10^{-6} $).
\end{corollary}

\begin{IEEEproof}
Let the state vector of each plant obey Gaussian distribution $ \mathbb{N}(\myvec{0},\sigma^2_\text{PS}\myvec{I}_N) $, then we have
\begin{align}
R^\text{S}_k\left( d_k \right) &= \dfrac{1}{2}\log_2\left( \dfrac{\sigma^2_\text{PS}}{d_k} \right), \\
W^\text{S}_k &= \dfrac{1}{2}(\log_2 \myexp)^2.
\end{align}
Based on~\eqref{approxV} in Appendix~\ref{app:theo1}, this corollary is the direct application of Theorem~\ref{theo1} and~\eqref{setradoff}.
\end{IEEEproof}

\subsection{LQR Cost of Control Phase}
By jointly considering the control efficiency and the SE in the control phase, the LQR cost is investigated for the optimization framework. The theorem about the LQR cost is shown as follows.

\begin{theorem}
\label{theo2}
For the single-cell networked control system with URLLCs, during the control phase, the LQR cost function about the transmission latency, reliability and power is
\begin{align}
\label{LQRcost}
c_k\left( p_k^\text{C},L^\text{C},\epsilon_k \right) = \dfrac{N_k Z_k\left( \myvec{w}_k \right) \left| \det \left( \myvec{M}_k \right) \right|^\frac{1}{N_k}}{\exp\left( \dfrac{2}{N_k} \Omega_k \ln 2 \right) - 1} + c^\text{min}_k, \forall k,
\end{align}
where $ c^\text{min}_k = \tr(\myvec{\Sigma}_{\myvec{w},k} \myvec{S}_k) $ and
\begin{align}
\label{Omega}
\Omega_k &= L^\text{C}B\log_2\left( 1+\dfrac{p_k^\text{C}\chi_k\beta_k}{\sigma^2_\text{C}} \right) - \sqrt{L^\text{C}B}Q^{-1}\left( \epsilon_k \right) \log_2 \myexp \notag \\
&\quad - \log_2 \left| \det \left( \myvec{A}_k \right) \right|.
\end{align}
\end{theorem}

\begin{IEEEproof}
See Appendix~\ref{app:theo2}.
\end{IEEEproof}

\subsection{Energy Consumption of Sensing Phase and Control Phase}
Energy consumption is another crucial aspect for the optimization framework. In the sensing phase, the energy consumption at the $ k $-th plant can be written as
\begin{align}
E^\text{S}_k\left( p_k^\text{S},L^\text{S} \right) = \dfrac{p^\text{S}_{k} L^\text{S}}{\mu^\text{S}_k} + P^\text{S}_k L^\text{S}, \forall k,
\end{align}
where $ \mu^\text{S}_k $ is the efficiency of the power amplifier at the $ k $-th plant, and $ P^\text{S}_k $ denotes the circuit power to operate each plant. Therefore, the total energy consumption during the sensing phase is given by
\begin{align}
E^\text{S}\left( \myvec{p}^\text{S},L^\text{S} \right) = \sum^{K}_{k=1} E^\text{S}_k\left( p_k^\text{S},L^\text{S} \right) = L^\text{S} \sum^{K}_{k=1} \left( \dfrac{p^\text{S}_{k}}{\mu^\text{S}_k} + P^\text{S}_k \right).
\end{align}
In the control phase, the energy consumption at the control center for the $ k $-th plant can be expressed as
\begin{align}
E^\text{C}_k\left( p_k^\text{C},L^\text{C} \right) = \dfrac{p^\text{C}_{k} L^\text{C}}{\mu^\text{C}} + \dfrac{P^\text{C} L^\text{C}}{K}, \forall k,
\end{align}
where $ \mu^\text{C} $ is the efficiency of the power amplifier at the control center, while $ P^\text{C} $ represents the circuit power to operate the control center. Similarly, the total energy consumption during the control phase is given by
\begin{align}
E^\text{C}\left( \myvec{p}^\text{C},L^\text{C} \right) = \sum^{K}_{k=1} E^\text{C}_k\left( p_k^\text{C},L^\text{C} \right) = L^\text{C} \left( \sum^{K}_{k=1}\dfrac{p^\text{C}_{k}}{\mu^\text{C}} + P^\text{C} \right).
\end{align}

\subsection{Novel Performance Metric for Networked Control System with URLLCs}
So far, we derive two vital components for the optimization framework, namely the LQR cost and the energy consumption. For the networked control system with URLLCs, our first optimization objective is to minimize the LQR cost of the control phase. At the same time, our second optimization objective is to minimize the round-trip energy consumption. To this end, in order to optimize these two objectives together, we develop a novel performance metric for the optimization framework, i.e., the energy-to-control efficiency (ECE). Two types of ECE can be written as
\begin{align}
\label{GECE}
\eta^\text{GECE} &= \dfrac{E^\text{GECE}_\text{max}-E^\text{S}\left( \myvec{p}^\text{S},L^\text{S} \right)-E^\text{C}\left( \myvec{p}^\text{C},L^\text{C} \right)}{\sum^{K}_{k=1} c_k\left( p_k^\text{C},L^\text{C},\epsilon_k \right)}, \\
\label{FECE}
\eta^\text{FECE} &= \min_k\left\lbrace \dfrac{E^\text{FECE}_\text{max}-E^\text{S}_k\left( p_k^\text{S},L^\text{S} \right)-E^\text{C}_k\left( p_k^\text{C},L^\text{C} \right)}{c_k\left( p_k^\text{C},L^\text{C},\epsilon_k \right)} \right\rbrace,
\end{align}
where $ E^\text{GECE}_\text{max} $ and $ E^\text{FECE}_\text{max} $ denote the round-trip maximum energy consumption, and they are given by
\begin{align}
E^\text{GECE}_\text{max} &= L_\text{max}\left[ \dfrac{KP_\text{S,max}}{\min_k\left\lbrace \mu^\text{S}_k \right\rbrace} + K\max_k\left\lbrace P^\text{S}_k \right\rbrace + \dfrac{P_\text{C,max}}{\mu^\text{C}} + P^\text{C} \right], \\
E^\text{FECE}_\text{max} &= L_\text{max}\left[ \dfrac{P_\text{S,max}}{\min_k\left\lbrace \mu^\text{S}_k \right\rbrace} + \max_k\left\lbrace P^\text{S}_k \right\rbrace + \dfrac{P_\text{C,max}}{\mu^\text{C}} + \dfrac{P^\text{C}}{K} \right].
\end{align}
$ P_\text{S,max} $ and $ P_\text{C,max} $ are the maximum transmission power during two phases, and $ L_\text{max} $ denotes the requirement of latency. Based on~\eqref{GECE} and \eqref{FECE}, one can find that~\eqref{GECE} focuses on the global performance of ECE, while~\eqref{FECE} focuses on the fairness of ECE. Although the physical meaning of ECE is undefined, ECE is a reasonable and valid performance utility for the networked control system with URLLCs. Next, let us show the \textit{rationality and validity} of the global ECE (GECE) and the fair ECE (FECE).

\begin{theorem}
\label{theo3}
For the noise of the control process with $ Z_k(\myvec{w}_k) > 0 $, $ \eta^\text{GECE} $ is a \textit{pseudo-concave} function of $ \myvec{p} = [\myvec{p}^\text{S},\myvec{p}^\text{C}] $ or $ \myvec{\tau} = [L^\text{S},L^\text{C}] $, and $ \eta^\text{FECE} $ is a \textit{quasi-concave} function of $ \myvec{p} = [\myvec{p}^\text{S},\myvec{p}^\text{C}] $ or $ \myvec{\tau} = [L^\text{S},L^\text{C}] $.
\end{theorem}

\begin{IEEEproof}
See Appendix~\ref{app:theo3}.
\end{IEEEproof}

\begin{corollary}
\label{coro2}
For the single-cell networked control system with URLLCs, maximizing $ \eta^\text{GECE} $ or $ \eta^\text{FECE} $ means that it is an efficient and optimal system design.
\end{corollary}

\begin{IEEEproof}
Since the local maximum is also the global maximum both for pseudo-concave and quasi-concave functions, Theorem~\ref{theo3} directly leads to this corollary.
\end{IEEEproof}

In conclusion, the proposed ECE can guide the design of the networked control system.

\section{Performance Optimization of Networked Control System with URLLCs}
\label{sec:Optimization}

Based on the optimization framework proposed, we optimize the performance of the single-cell networked control system in this section. We first formulate a general and complex max-min joint optimization problem. Then, a series of radio resource allocation algorithms are proposed to optimally solve the formulated problem by reasonable complexity.

\subsection{Max-Min Joint Optimization Problem Statement}

According to Theorem~\ref{theo3} and Corollary~\ref{coro2}, due to the pseudo-concavity, the Karush-Kuhn-Tucker (KKT) conditions are sufficient and necessary for maximizing $ \eta^\text{GECE} $ when we separately optimize the power allocation or the latency allocation. To this end, we formulate a more \textit{general and complex} max-min joint optimization problem in this section. In addition, since the expected goal of the networked control system is to let each plant stabilize, maximizing the minimum ECE $ \eta^\text{FECE} $ among all plants is more in line with actual needs.

\begin{problem}[Max-Min Joint Optimization of Transmission Power and Latency]
\label{pro1}
Given the location information of all plants and the URLLC requirements $ L_\text{max} $, $ \delta_k $ and $ \epsilon_k $, the joint optimization of the transmission power and latency is formulated as
\begin{subequations}
\label{problem1}
\begin{align}
\max_{\substack{\myvec{p}^\text{S}, \myvec{p}^\text{C} \\ L^\text{S}, L^\text{C}}}~ & \min_k \left\lbrace \dfrac{E^\text{FECE}_\text{max}-E^\text{S}_k\left( p_k^\text{S},L^\text{S} \right)-E^\text{C}_k\left( p_k^\text{C},L^\text{C} \right)}{c_k\left( p_k^\text{C},L^\text{C},\epsilon_k \right)} \right\rbrace \\
\label{pcon1}
\sbjto~ & L^\text{S}B\log_2\left( 1+\dfrac{p^\text{S}_{k}\chi_k\beta_k}{\sigma^2_\text{S}} \right) \geqslant \dfrac{N_k}{2}\log_2\left( \dfrac{\sigma^2_\text{PS}}{d_k} \right) \notag \\
& \qquad \qquad + \sqrt{L^\text{S}B + \dfrac{N_k}{2}} Q^{-1}\left( \delta_k \right) \log_2 \myexp, \forall k, \\
\label{pcon2}
& p^\text{S}_k \leqslant P_\text{S,max}, \forall k, \\
\label{pcon3}
& \sum^{K}_{k=1} p^\text{C}_{k} \leqslant P_\text{C,max}, \\
\label{pcon4}
& p^\text{C}_k \geqslant 0, \forall k, \\
\label{pcon5}
& p^\text{S}_k \geqslant 0, \forall k, \\
\label{lcon1}
& L^\text{S} + L^\text{C} \leqslant L_\text{max}, \\
\label{lcon2}
& L^\text{C} \geqslant 0, \\
\label{lcon3}
& L^\text{S} \geqslant 0.
\end{align}
\end{subequations}
\end{problem}

\subsection{Solution to Problem~\ref{pro1}}
As shown in \eqref{problem1}, Problem~\ref{pro1} is non-concave when we jointly optimize the transmission power and latency. However, the following theorem illustrates that Problem~\ref{pro1} can still be solved optimally. To facilitate exposition, we write the following notations, i.e.,
\begin{align}
& \dfrac{E^\text{FECE}_\text{max}-E^\text{S}_k\left( p_k^\text{S},L^\text{S} \right)-E^\text{C}_k\left( p_k^\text{C},L^\text{C} \right)}{c_k\left( p_k^\text{C},L^\text{C},\epsilon_k \right)} \triangleq \dfrac{f_k\left( \myvec{p}, \myvec{\tau} \right)}{g_k\left( \myvec{p}, \myvec{\tau} \right)}, \forall k, \\
& F\left( \eta \right) \triangleq \max_{\myvec{p}, \myvec{\tau}}\min_{k} \left\lbrace f_k\left( \myvec{p}, \myvec{\tau} \right)-\eta g_k\left( \myvec{p}, \myvec{\tau} \right) \right\rbrace.
\end{align}

\begin{theorem}
\label{theo4}
Denote $ \mathcal{F} $ as the set of all feasible solutions. Let $ \eta^* = \min_{k}\left\lbrace f_k(\myvec{p}^*, \myvec{\tau}^*)/g_k(\myvec{p}^*, \myvec{\tau}^*) \right\rbrace $, then $ (\myvec{p}^*, \myvec{\tau}^*) $ is the optimal solution of Problem~\ref{pro1} if and only if
\begin{align}
\label{opt}
\left( \myvec{p}^*, \myvec{\tau}^* \right) = \arg\max_{\myvec{p}, \myvec{\tau}}\min_{k} \left\lbrace f_k\left( \myvec{p}, \myvec{\tau} \right)-\eta^* g_k\left( \myvec{p}, \myvec{\tau} \right) \right\rbrace.
\end{align}
\end{theorem}

\begin{IEEEproof}
The proof of Theorem~\ref{theo4} is divided into sufficiency and necessity. First of all, let $ (\myvec{p}^*, \myvec{\tau}^*) $ be the optimal solution of Problem~\ref{pro1}, then $ \forall (\myvec{p}, \myvec{\tau}) \in \mathcal{F} $, we obtain
\begin{align}
\eta^* = \min_{k}\left\lbrace \dfrac{f_k\left( \myvec{p}^*, \myvec{\tau}^* \right)}{g_k\left( \myvec{p}^*, \myvec{\tau}^* \right)} \right\rbrace \geqslant \min_{k}\left\lbrace \dfrac{f_k\left( \myvec{p}, \myvec{\tau} \right)}{g_k\left( \myvec{p}, \myvec{\tau} \right)} \right\rbrace.
\end{align}
Therefore,
\begin{align}
\min_{k}\left\lbrace f_k\left( \myvec{p}, \myvec{\tau} \right) - \eta^* g_k\left( \myvec{p}, \myvec{\tau} \right) \right\rbrace &\leqslant 0, \\
\min_{k}\left\lbrace f_k\left( \myvec{p}^*, \myvec{\tau}^* \right) - \eta^* g_k\left( \myvec{p}^*, \myvec{\tau}^* \right) \right\rbrace &= 0,
\end{align}
which yields~\eqref{opt}.

On the other hand, let~\eqref{opt} hold, then $ \forall (\myvec{p}, \myvec{\tau}) \in \mathcal{F} $, we have
\begin{align}
\label{necc}
&\mathrel{\phantom{\leqslant}} \min_{k}\left\lbrace f_k\left( \myvec{p}, \myvec{\tau} \right) - \eta^* g_k\left( \myvec{p}, \myvec{\tau} \right) \right\rbrace \notag \\
& \leqslant \min_{k}\left\lbrace f_k\left( \myvec{p}^*, \myvec{\tau}^* \right) - \eta^* g_k\left( \myvec{p}^*, \myvec{\tau}^* \right) \right\rbrace = F\left( \eta^* \right) \overset{\text{(a)}}{=} 0,
\end{align}
where (a) follows from Lemma~\ref{lemma3}.

\begin{lemma}[\cite{YangTVT2019}]
\label{lemma3}
$ \forall \myvec{a} \in \mathcal{F} $, we have $ F(\eta_{\myvec{a}}) \geqslant 0 $ with $ \eta_{\myvec{a}} = \min_{k}\left\lbrace f_k(\myvec{a})/g_k(\myvec{a}) \right\rbrace $, and the equality holds when $ \myvec{a} = \arg \max\limits_{\myvec{a}}\min\limits_{k} \left\lbrace f_k(\myvec{a}) - \eta_{\myvec{a}} g_k(\myvec{a}) \right\rbrace $.
\end{lemma}

Based on \eqref{necc}, one can conclude that
\begin{align}
\eta^* &\geqslant \min_{k}\left\lbrace \dfrac{f_k\left( \myvec{p}, \myvec{\tau} \right)}{g_k\left( \myvec{p}, \myvec{\tau} \right)} \right\rbrace, \\
\eta^* &= \min_{k}\left\lbrace \dfrac{f_k\left( \myvec{p}^*, \myvec{\tau}^* \right)}{g_k\left( \myvec{p}^*, \myvec{\tau}^* \right)} \right\rbrace.
\end{align}
Thus, $ (\myvec{p}^*, \myvec{\tau}^*) $ is the optimal solution of Problem~\ref{pro1}.
\end{IEEEproof}

\subsection{Radio Resource Allocation Algorithm for Max-Min Joint Optimization}

\begin{figure}[!t]
\begin{MYalgorithmic}
\algcaption{Radio resource allocation algorithm for max-min joint optimization.}
\label{alg1}
\begin{algorithmic}[5]
	\renewcommand{\algorithmicrequire}{\textbf{Initialization:}}
	\Require
	\State \labelitemi~The location information of all plants.
	\State \labelitemi~The URLLC requirements, namely $ L_\text{max} $, $ \delta_k $ and $ \epsilon_k, \forall k $.
	\State \labelitemi~The distortion requirement of the sensing phase, namely $ \frac{\sigma^2_\text{PS}}{d_k}, \forall k $.
	\State \labelitemi~Total power $ P_\text{S,max} $ and $ P_\text{C,max} $.
	\State \labelitemi~Iterative index $ i=0 $ and maximum iterative tolerance $ \zeta_1 > 0 $.
	\State \labelitemi~Initial values of $ \eta_0 = \min_{k} \left\lbrace \frac{E^\text{FECE}_\text{max}}{2 c^\text{min}_k} \right\rbrace $ and $ F(\eta_{-1}) > \zeta_1 $.
	\renewcommand{\algorithmicrequire}{\textbf{Iterative procedure:}}
	\Require
	\While{$ F(\eta_{i-1}) > \zeta_1 $} % no abs(\cdot) due to Lemma~\ref{lemma3}.
	\State {\algitem\label{sovsp}}~Based on Algorithm~\ref{alg2}, solve the resource allocation $ \myvec{p}_i $ and $ \myvec{\tau}_i $ of Sub-Problem~\ref{subpro2} with the given $ \eta_i $.
	\State \algitem~Calculate the value of auxiliary function $ F(\eta_{i}) = \min_{k} \left\lbrace f_k(\myvec{p}_i, \myvec{\tau}_i)-\eta_i g_k(\myvec{p}_i, \myvec{\tau}_i) \right\rbrace $.
	\State {\algitem\label{rule}}~Update $ \eta_{i+1} = \min_{k}\left\lbrace f_k(\myvec{p}_i, \myvec{\tau}_i) / g_k(\myvec{p}_i, \myvec{\tau}_i) \right\rbrace $, $ \myvec{p}_{i+1} = \myvec{p}_{i} $ and $ \myvec{\tau}_{i+1} = \myvec{\tau}_{i} $.
	\State \algitem~Set $ i=i+1 $.
	\EndWhile
	\State \algitem~Output the optimal solution $ (\myvec{p}^*, \myvec{\tau}^*) = (\myvec{p}_{i+1}, \myvec{\tau}_{i+1}) $ and the optimal FECE $ \eta^* $.
\end{algorithmic}
\end{MYalgorithmic}
\end{figure}

Based on Theorem~\ref{theo4}, we find that $ F(\eta^*) = 0 $ with $ \eta^* = \min_{k}\left\lbrace f_k(\myvec{p}^*, \myvec{\tau}^*)/g_k(\myvec{p}^*, \myvec{\tau}^*) \right\rbrace $ when $ (\myvec{p}^*, \myvec{\tau}^*) $ is the optimal solution of Problem~\ref{pro1}. Furthermore, because $ F(\eta) $ is strictly monotonically decreasing on $ \eta $, and we have $ \lim_{\eta \rightarrow -\infty}F(\eta) = +\infty, \lim_{\eta \rightarrow +\infty}F(\eta) = -\infty $. Consequently, solving Problem~\ref{pro1} is equivalent to finding the unique zero root of $ F(\eta) $. In this section, a Dinkelbach-based iterative procedure described by Algorithm~\ref{alg1} is put forward to solve Problem~\ref{pro1}. The convergence and optimality of Algorithm~\ref{alg1} are guaranteed by the following theorem.

\begin{theorem}
\label{theo5}
Algorithm~\ref{alg1} must converge to the optimal solution of Problem~\ref{pro1}.
\end{theorem}

\begin{IEEEproof}
Step~\ref{rule} of Algorithm~\ref{alg1} gives rise to
\begin{align}
\label{cov}
F\left( \eta_{i} \right) &= \min_{k} \left\lbrace f_k\left( \myvec{p}_i, \myvec{\tau}_i \right)-\eta_i g_k\left( \myvec{p}_i, \myvec{\tau}_i \right) \right\rbrace \notag \\
&\overset{\text{(a)}}{=} \min_{k} \left\lbrace \left( \eta_{i+1}-\eta_i \right) g_k\left( \myvec{p}_i, \myvec{\tau}_i \right) \right\rbrace \notag \\
&\overset{\text{(b)}}{\geqslant} 0.
\end{align}
(a) follows from
\begin{align}
&\mathrel{\phantom{\Rightarrow}} \dfrac{f_k\left( \myvec{p}_i, \myvec{\tau}_i \right)}{g_k\left( \myvec{p}_i, \myvec{\tau}_i \right)} \geqslant \eta_{i+1} = \min_{k}\left\lbrace \dfrac{f_k\left( \myvec{p}_i, \myvec{\tau}_i \right)}{g_k\left( \myvec{p}_i, \myvec{\tau}_i \right)} \right\rbrace, \forall k \notag \\
&\Rightarrow f_k\left( \myvec{p}_i, \myvec{\tau}_i \right) \geqslant \eta_{i+1}g_k\left( \myvec{p}_i, \myvec{\tau}_i \right),
\end{align}
where the equality holds when $ k = \arg\min_{k}\left\lbrace f_k(\myvec{p}_i, \myvec{\tau}_i) / g_k(\myvec{p}_i, \myvec{\tau}_i) \right\rbrace $. While (b) follows from Lemma~\ref{lemma3}. Therefore, \eqref{cov} means that $ \eta_{i+1}>\eta_i $ always holds when Algorithm~\ref{alg1} does not achieve the convergence. Based on the property of monotonically decreasing on $ F(\eta) $, this completes the proof of the convergence ($ F(\eta) \rightarrow 0 $).

On the other hand, the optimality can be proved by contradiction. Let $ \tilde{\eta} = \min_{k}\left\lbrace f_k(\tilde{\myvec{p}}, \tilde{\myvec{\tau}}) / g_k(\tilde{\myvec{p}}, \tilde{\myvec{\tau}}) \right\rbrace $ be a feasible solution which satisfies $ \tilde{\eta} < \eta^* $. With the aid of the monotonicity and convergence of $ F(\eta) $, we obtain $ F(\tilde{\eta})>F(\eta^*) $ and $ F(\tilde{\eta})=0 $. By contrast, based on Theorem~\ref{theo3}, $ F(\eta^*)=0 $, which leads to a contradiction.
\end{IEEEproof}

\subsection{Radio Resource Allocation Algorithm for Max-Min Sub-Problem}
As illustrated in Theorem~\ref{theo5}, Algorithm~\ref{alg1} can achieve the optimal solution of Problem~\ref{pro1}. In Algorithm~\ref{alg1}, Step~\ref{sovsp} gives rise to the following original sub-problem.

\begin{subproblem}[Original Sub-Problem]
\label{subpro1}
Given $ \eta_i $, the original max-min sub-problem is formulated as
\begin{subequations}
\label{subproblem1}
\begin{align}
\label{subproblem1obj}
\max_{\myvec{p}, \myvec{\tau}}~ & \min_{k} \left\lbrace f_k\left( \myvec{p}, \myvec{\tau} \right)-\eta_i g_k\left( \myvec{p}, \myvec{\tau} \right) \right\rbrace \\
\sbjto~ & \eqref{pcon1} - \eqref{lcon3}.
\end{align}
\end{subequations}
\end{subproblem}

Since the condition $ x \geqslant \ln(\frac{\myexp}{\myexp-1}) \approx 0.4587 $ always holds for practical engineering implementations, then we have 
the inequality $ \exp(x)-1 \geqslant \exp(x-1) $. Therefore, the upper bound of $ g_k(\myvec{p}, \myvec{\tau}) $ is adopted to further reduce the complexity, i.e.,
\begin{align}
\label{upperg}
g_k\left( \myvec{p}, \myvec{\tau} \right) \leqslant \tilde{g}_k\left( \myvec{p}, \myvec{\tau} \right) = \dfrac{N_k Z_k\left( \myvec{w}_k \right) \left| \det \left( \myvec{M}_k \right) \right|^\frac{1}{N_k}}{\exp\left( \dfrac{2}{N_k} \Omega_k \ln 2 -1 \right)} + c^\text{min}_k, \forall k,
\end{align}
where $ \Omega_k $ is given by~\eqref{Omega}. Clearly, \eqref{upperg} gives rise to the lower bound of \eqref{subproblem1obj}. Then, based on the standard epigraph form, the transformed sub-problem is established as follows.

\begin{subproblem}[Transformed Sub-Problem]
\label{subpro2}
Given $ \eta_i $, the transformed sub-problem is formulated as
\begin{subequations}
\label{subproblem2}
\begin{align}
\max_{\myvec{p}, \myvec{\tau}, \psi}~ & \left\lbrace \psi \right\rbrace \\
\sbjto~ & f_k\left( \myvec{p}, \myvec{\tau} \right)-\eta_i \tilde{g}_k\left( \myvec{p}, \myvec{\tau} \right) \geqslant \psi, \forall k, \\
& - \infty \leqslant \psi \leqslant \min_{k} \left\lbrace E^\text{FECE}_\text{max}-\eta_i c^\text{min}_k \right\rbrace, \\
& \eqref{pcon1} - \eqref{lcon3}.
\end{align}
\end{subequations}
\end{subproblem}

\subsubsection{Radio Resource Allocation Algorithm for Sub-Problem~\ref{subpro2}}

\begin{figure}[!t]
\begin{MYalgorithmic}
\algcaption{Radio resource allocation algorithm for Sub-Problem~\ref{subpro2}.}
\label{alg2}
\begin{algorithmic}[5]
	\renewcommand{\algorithmicrequire}{\textbf{Initialization:}}
	\Require
	\State \labelitemi~Iterative index $ j=0 $ and maximum iterative tolerance $ \zeta_2 > 0 $.
	\State \labelitemi~Initial value of $ \myvec{\tau}_i^\text{I} = \frac{L_\text{max}}{2} \cdot \myvec{1} $.
	\renewcommand{\algorithmicrequire}{\textbf{Iterative procedure:}}
	\Require
	\State \algitem~Set $ \psi^\text{min}_0 = \min_{k} \left\lbrace - \eta_i c^\text{min}_k \right\rbrace $ and $ \psi^\text{max}_0 = \min_{k} \left\lbrace E^\text{FECE}_\text{max}-\eta_i c^\text{min}_k \right\rbrace $.
	\While{$ \left| \psi^\text{max}_j - \psi^\text{min}_j \right| > \zeta_2 $}
	\State \algitem~Let $ \psi_j = \frac{1}{2}\left( \psi^\text{min}_j+\psi^\text{max}_j \right) $.
	\State {\algitem\label{sovfesp}}~Solve the feasibility problem \eqref{problemfesp} with the given $ \eta_i $, $ \psi_j $ and $ \myvec{\tau}_i^\text{I} $ to obtain the solution of $ \myvec{p}_i^\text{F} $.
	\State {\algitem\label{sovfesl}}~Solve the feasibility problem or \eqref{problemfesl} with the given $ \eta_i $, $ \psi_j $ and $ \myvec{p}_i^\text{F} $ to obtain the solution of $ \myvec{\tau}_i^\text{F} $.
	\If{$ \myvec{p}_i^\text{F} $ and $ \myvec{\tau}_i^\text{F} $ are \textbf{\textit{all}} feasible}~Let $ \psi^\text{min}_{j+1} = \psi_{j} $, $ \psi^\text{max}_{j+1} = \psi^\text{max}_{j} $ and $ \myvec{\tau}_i^\text{I} = \myvec{\tau}_i^\text{F} $.
	\Else~Let $ \psi^\text{max}_{j+1} = \psi_{j} $ and $ \psi^\text{min}_{j+1} = \psi^\text{min}_{j} $.
	\EndIf
	\State \algitem~Set $ j=j+1 $.
	\EndWhile
	\State \algitem~Output the solution $ (\myvec{p}_{i}, \myvec{\tau}_{i}) = (\myvec{p}_i^\text{F}, \myvec{\tau}_i^\text{F}) $.
\end{algorithmic}
\end{MYalgorithmic}
\end{figure}

Since $ \psi $ has the specific lower and upper bounds in practice, a binary search-based iterative procedure described by Algorithm~\ref{alg2} is proposed to solve Sub-Problem~\ref{subpro2}. Due to the coupled variables of $ \myvec{p} $ and $ \myvec{\tau} $, the transformed sub-problem is still non-concave. Consequently, the power allocation and the latency allocation are optimized separately. Moreover, it is noted that the FECE $ \eta^\text{FECE} $ may not be optimal when the equalities of \eqref{pcon2}, \eqref{pcon3} and \eqref{lcon1} hold. Thus, we cannot let $ \myvec{p}^\text{S} = P_\text{S,max} \cdot \myvec{1} $ or $ L^\text{S} = L_\text{max} - L^\text{C} $ to reduce the complexity. Finally, the convergence and optimality of Algorithm~\ref{alg2} can be guaranteed by the binary search.

\subsubsection{Power Allocation}
Firstly, let us solve the constraint \eqref{pcon1} with the given $ L^\text{S} $, then we obtain
\begin{align}
p^\text{S}_{k} \geqslant \dfrac{\sigma^2_\text{S}}{\chi_k\beta_k} \left[ \exp\left( \Phi_k \right) - 1 \right], \forall k,
\end{align}
where
\begin{align}
\Phi_k = \dfrac{N_k\log_2\left( \dfrac{\sigma^2_\text{PS}}{d_k} \right)}{2L^\text{S}B\log_2 \myexp} + \dfrac{\sqrt{L^\text{S}B + \dfrac{N_k}{2}} Q^{-1}\left( \delta_k \right)}{L^\text{S}B}.
\end{align}
Then, the feasibility problem in Step~\ref{sovfesp} of Algorithm~\ref{alg2} can be formulated as
\begin{subequations}
\label{problemfesp}
\begin{align}
\textbf{\text{FP-P:}}\quad \min_{\myvec{b}^\text{S}, b^\text{C}, \myvec{p}}~ & \left\lbrace b^\text{C}+\sum_{k=1}^K b^\text{S}_k \right\rbrace \\
\sbjto~ & 0 \leqslant b^\text{S}_k \leqslant 1, \forall k, \\
& \dfrac{\sigma^2_\text{S}}{\chi_k\beta_k} \left[ \exp\left( \Phi_k \right) - 1 \right] \leqslant p^\text{S}_k \leqslant b^\text{S}_k \cdot P_\text{S,max}, \forall k, \\
& 0 \leqslant b^\text{C} \leqslant 1, \\
& \sum^{K}_{k=1} p^\text{C}_{k} \leqslant b^\text{C} \cdot P_\text{C,max}, \\
& p^\text{C}_k \geqslant 0, \forall k, \\
& f_k\left( \myvec{p} \right)-\eta_i \tilde{g}_k\left( \myvec{p} \right) \geqslant \psi, \forall k,
\end{align}
\end{subequations}
where $ \myvec{b}^\text{S} $ and $ b^\text{C} $ denote the indicators of feasibility for the power allocation.

\subsubsection{Latency Allocation}
According to the discriminant of the quadratic equation, $ L^\text{S} $ has the unique positive solution for each plant. Analogously, based on the constraint \eqref{pcon1}, we can solve one of the lower bounds of $ L^\text{S} $ with the given $ p^\text{S}_{k} $, i.e.,
\begin{align}
\label{lconsov}
L^\text{S} \geqslant \dfrac{2q_{0,k}q_{1,k}+B+\sqrt{\left( 4Bq_{0,k}q_{1,k}+2N_k q_{0,k}^2+B^2 \right)}}{2q_{0,k}^2}, \forall k,
\end{align}
where
\begin{align}
q_{0,k} &= \dfrac{B\log_2\left( 1+\dfrac{p^\text{S}_{k}\chi_k\beta_k}{\sigma^2_\text{S}} \right)}{Q^{-1}\left( \delta_k \right) \log_2 \myexp}, \\
q_{1,k} &= \dfrac{N_k \log_2\left( \dfrac{\sigma^2_\text{PS}}{d_k} \right)}{2Q^{-1}\left( \delta_k \right) \log_2 \myexp}.
\end{align}
Then, the feasibility problem in Step~\ref{sovfesl} of Algorithm~\ref{alg2} is
\begin{subequations}
\label{problemfesl}
\begin{align}
\textbf{\text{FP-L:}}\quad \min_{b, \myvec{\tau}}~ & \left\lbrace b \right\rbrace \\
\sbjto~ & 0 \leqslant b \leqslant 1, \\
& L^\text{S} + L^\text{C} \leqslant b \cdot L_\text{max}, \\
& L^\text{S} \geqslant \max_{k} \left\lbrace \text{RHS of~}\eqref{lconsov} \right\rbrace, \\
& L^\text{C} \geqslant 0, \\
& f_k\left( \myvec{\tau} \right)-\eta_i \tilde{g}_k\left( \myvec{\tau} \right) \geqslant \psi, \forall k,
\end{align}
\end{subequations}
where $ b $ is the indicators of feasibility for latency allocation.

To reduce the complexity, the \textsf{CVX} package with the solver \textsf{SDPT3}~\cite{cvx,gb08} is used for solving the feasibility problems in this paper.

So far, we optimally solve Problem~\ref{pro1}. Subsequently, we discuss how to apply Algorithm~\ref{alg1} to jointly optimize the transmission power and latency for maximizing $ \eta^\text{GECE} $.

\begin{remark}[Joint Optimization of Maximizing $ \eta^\text{GECE} $]
Through analysis and comparison, we find that all proofs on Algorithm~\ref{alg1} still hold for maximizing $ \eta^\text{GECE} $. Furthermore, although the KKT conditions are sufficient and necessary for various maximized sub-problems of the power allocation or the latency allocation, the closed-form solutions may not be achieved. At this point, the \textsf{CVX} package with the solver \textsf{SDPT3} or the standard sub-gradient method~\cite{Boydbook} can be used to obtain the optimal solutions. In a word, the problem of maximizing $ \eta^\text{GECE} $ is easier than that of maximizing $ \eta^\text{FECE} $.
\end{remark}

\section{Simulation Results}
\label{sec:Simulation}

\subsection{Simulation Setup}

\begin{table}[!t]
	\setlength{\extrarowheight}{1pt}
	\centering
	\caption{Simulation Parameters}
	\begin{tabular}{ l | r }
		\hline
		\textbf{Parameter} & \textbf{Value} \\
		\hline
		\hline
		Inner radius $ r_\text{I} $ & 10~\myunit{m} \\
		\hline
		Outer radius $ r_\text{O} $ & 50~\myunit{m} \\
		\hline
		Channel constant $ \theta $ & $ 10^{-3} $ \\ 
		\hline
		SF standard variance $ \sigma_\text{SF} $ & 8~\myunit{dB} \\
		\hline
		Path loss exponent $ \alpha $ & 3 \\
		\hline
		Estimation accuracy $ \chi_k, \forall k $ & 0.8 \\
		\hline
		Noise power spectrum density $ N^\text{S}_0 = N^\text{C}_0 $ & -130~\myunit{dBm/Hz}~\cite{Zhao2016} \\
		\hline
		System coherence bandwidth $ B $ & 500~\myunit{kHz} \\
		\hline
		Maximum round-trip latency $ L_\text{max} $ & 1~\myunit{ms} \\
		\hline
		Reliability $ \delta_k = \epsilon_k, \forall k  $ & $ 10^{-5} $ \\
		\hline
		\makecell*[l]{Maximum transmission power \\ at plant $ P_\text{S,max} $} & 1~\myunit{W} \\
		\hline
		\makecell*[l]{Maximum transmission power \\ at control center $ P_\text{C,max} $} & 10~\myunit{W} \\
		\hline
		Circuit power at plant $ P^\text{S}_k, \forall k $ & 2~\myunit{W} \\
		\hline
		Circuit power at control center $ P^\text{C} $ & $ 10^3 $~\myunit{W}~\cite{Emil2017} \\
		\hline
		Efficiency of power amplifier $ \mu^\text{C} = \mu^\text{S}_k, \forall k $ & 0.2 \\
		\hline
		The number of plant states $ N_k, \forall k $ & 100 \\
		\hline
		Minimum LQR cost $ c^\text{min}_k, \forall k $ & $ 10^{-1} $ \\
		\hline
	\end{tabular}
	\label{table_sim}
\end{table}

In the simulations, according to the representative value of channel delay spread $ T_\text{D}=1 $~\textmu\myunit{s}, the system coherence bandwidth can be calculated as $ B = 1/(2T_\text{D})=500 $~\myunit{kHz}~\cite{Tsebook}. Therefore, the noise power during two phases is given by $ \sigma^2_\text{S} = N^\text{S}_0 B $ and $ \sigma^2_\text{C} = N^\text{C}_0 B $, where $ N^\text{S}_0 $ and $ N^\text{C}_0 $ are the noise power spectral density. The URLLC requirements are set as $ L_\text{max} = 1 $~\myunit{ms} and $ \delta_k = \epsilon_k = 10^{-5} $, respectively~\cite{3gpp913}. The large-scale fading is given by $ \beta_k = \theta \xi_k d_k^{-\alpha} $, where $ \theta $ is a constant related to the antenna gain and carrier frequency, $ \xi_k $ is the shadow fading (SF) variable with $ 10\log_{10}\xi_k \sim \mathbb{N}(0,\sigma^2_\text{SF}) $, $ \alpha $ is the path loss exponent, and $ d_k = [x_k^2+y_k^2]^{1/2} $ is the distance between the $ k $-th plant and the control center. All plants are uniformly distributed on a disc with the inner radius $ r_\text{I} $ and the outer radius $ r_\text{O} $, then
\begin{align}
\begin{cases}
x_{k} = r_{k} \cos\phi_{k}, \\
y_{k} = r_{k} \sin\phi_{k},
\end{cases}
\end{align}
where $ r_{k} $ and $ \phi_{k} $ are the distributions defined by following
\begin{align}
\label{Cdist}
\begin{cases}
f_{r_{k}}\left( x \right) = \dfrac{2x}{r^2_\text{O}-r^2_\text{I}}, x \in \left[ r_\text{I}, r_\text{O} \right], \\
\phi_{k} \sim \text{unif}\left( 0, 2\pi \right).
\end{cases}
\end{align}
In addition, we consider an important control case in the simulations, i.e., the control goal is to minimize the mean square deviation between the plant state and the desired state $ \myvec{0} $. Then, we have $ \myvec{Q}_k = \myvec{I}_{N_k} $, $ \myvec{R} = \myvec{0} $ and $ \myvec{S}_k = \myvec{M}_k = \myvec{I}_{N_k} $. Let $ \myvec{w}_{t,k} $ be the i.i.d. AWGN with $ \myvec{w}_{t,k} \sim \mathbb{CN}(\myvec{0},\sigma^2_\text{PN} \myvec{I}_{N_k}) $, we obtain $ Z_k( \myvec{w}_k ) = \sigma^2_\text{PN} $, $ \det ( \myvec{M}_k ) = 1 $ and $ c^\text{min}_k = N_k \sigma^2_\text{PN} $. All detailed simulation parameters are listed in Table~\ref{table_sim}. Finally, ``PCSI'' and ``IPCSI'' represent perfect and imperfect channel state information, respectively.

%\begin{align}
%c_k\left( p_k^\text{C},L^\text{C},\epsilon_k \right) = \dfrac{N_k \sigma^2_\text{PN}}{\exp\left( \dfrac{2}{N_k} \Omega_k \ln 2 \right) - 1} + N_k \sigma^2_\text{PN}, \forall k,
%\end{align}
%where
%\begin{align}
%\Omega_k &= L^\text{C}B\log_2\left( 1+\dfrac{p_k^\text{C}\chi_k\beta_k}{\sigma^2_\text{C}} \right) - \sqrt{L^\text{C}B}Q^{-1}\left( \epsilon_k \right) \log_2 \myexp \notag \\
%&\quad - \log_2 \left| \det \left( \myvec{A}_k \right) \right|.
%\end{align}

\subsection{Tightness of Theorem~\ref{theo1}}

\begin{figure}[!t]
	\centering
	\includegraphics[scale=0.3]{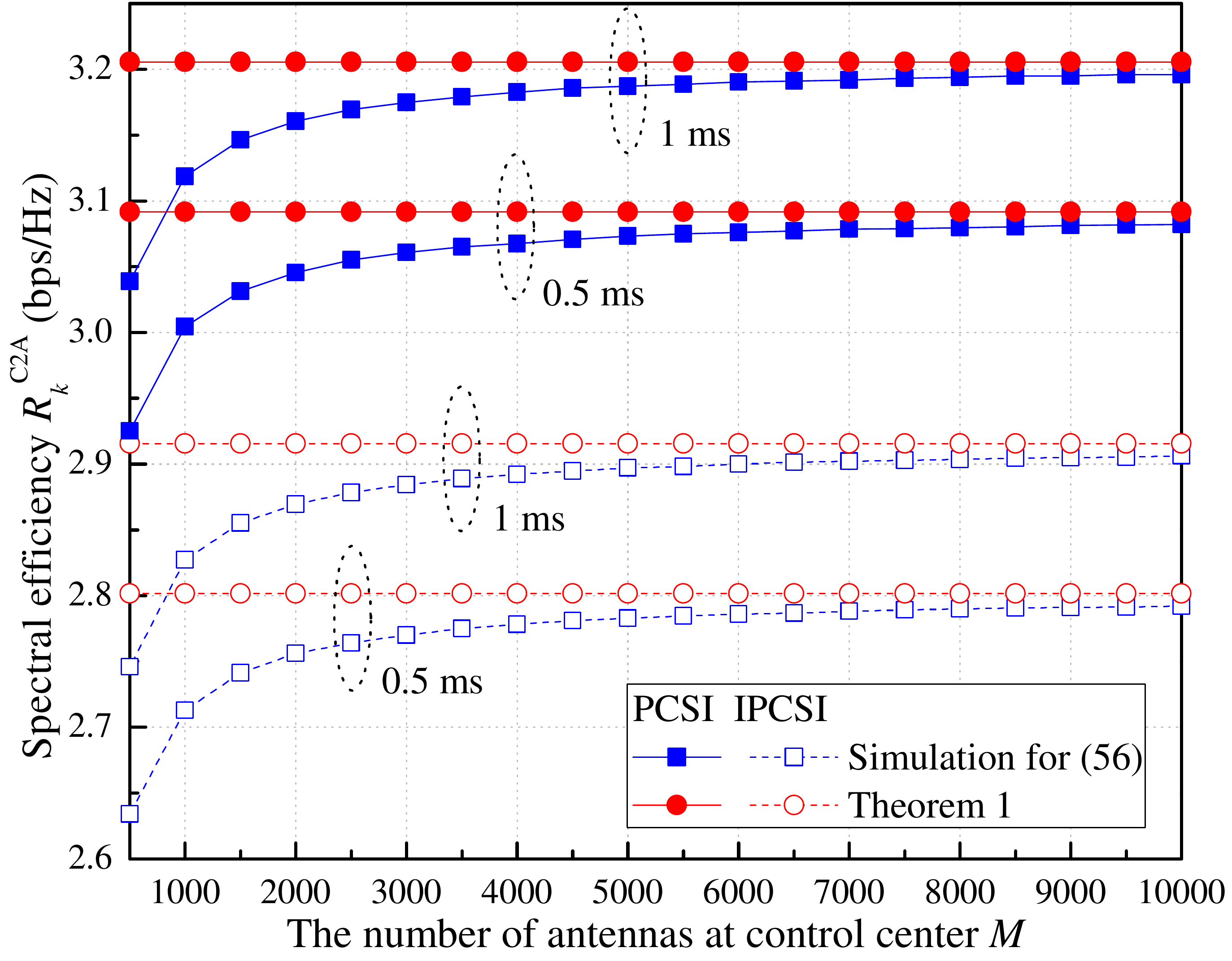}
	\caption{SE versus the number of antennas, with the fixed transmission power $ p^\text{C}_k = 0.5 $~\myunit{W}, $ \forall k $.}
	\label{fig_MvsSE}
\end{figure}

Let $ \myvec{a}_k \in \mathbb{C}^{M \times 1} $ denote an i.i.d. complex Gaussian random vector, namely $ \myvec{a}_k \mathop \sim \mathbb{CN}(\myvec{0},\myvec{I}_M) $, then we have the following distributions, i.e.,
\begin{align}
\mathsf{RV}_\text{B} &\triangleq \left| \dfrac{\myvec{a}_k^\text{H}}{\left\| \myvec{a}_k^\text{H} \right\|}\dfrac{\myvec{a}_i}{\left\| \myvec{a}_i \right\|} \right|^2 \sim \text{Beta}\left( 1,M-1 \right), \forall i \neq k, \\
\mathsf{RV}_\text{G} &\triangleq \left\| \myvec{a}_k \right\|^2 \sim \text{Gamma}\left( M,1 \right), \forall k.
\end{align}
Therefore, the instantaneous SINR of the $ k $-th plant during the control phase can be rewritten as
\begin{align}
\label{instantgammaC}
\gamma_k^\text{C} = \dfrac{p_k^\text{C}\chi_k\beta_k \mathsf{RV}_\text{G}}{\mathsf{RV}_\text{B} \mathsf{RV}_\text{G} \chi_k\beta_k \sum\limits_{\substack{i=1, \\ i \neq k}}^{K} p_i^\text{C} + \beta_k\left( 1-\chi_k \right) \sum\limits_{j=1}^{K} p_j^\text{C} + M \sigma^2_\text{C}}.
\end{align}
Based on \eqref{instantgammaC}, Fig.~\ref{fig_MvsSE} illustrates the SE versus the number of antennas during the control phase. For both PCSI and IPCSI, Fig.~\ref{fig_MvsSE} depicts that as the number of antennas increases, the ergodic SE of the control phase increases and tends to Theorem~\ref{theo1}. Hence, Theorem~\ref{theo1} is tight and valid for the large number of antennas $ M $ at the control center. Moreover, due to the channel estimation error, the case of IPCSI is always worse than the case of PCSI. Finally, Fig.~\ref{fig_MvsSE} shows that increasing the transmission latency deteriorates the SE performance, which means that the implementation of URLLCs comes at the cost of a reduced SE. The SE results of the sensing phase are similar to those of the control phase, thus they are omitted here due to space limitations. According to the tightness of Theorem~\ref{theo1}, the other simulation results will be discussed in the following subsections.

%\begin{align}
%\gamma_k^\text{S} = \dfrac{p^\text{S}_{k}\chi_k\beta_k \mathsf{RV}_\text{G}}{\mathsf{RV}_\text{B} \mathsf{RV}_\text{G} \sum\limits_{\substack{i=1, \\ i \neq k}}^{K} p^\text{S}_{i}\chi_i \beta_i + \sum\limits_{j=1}^{K} p^\text{S}_{j} \beta_j \left( 1-\chi_j \right) + M \sigma^2_\text{S}}.
%\end{align}

\subsection{Tradeoff Among LQR Cost, Transmission Latency and Transmission Reliability}

\begin{figure}[!t]
	\centering
	\includegraphics[scale=0.3]{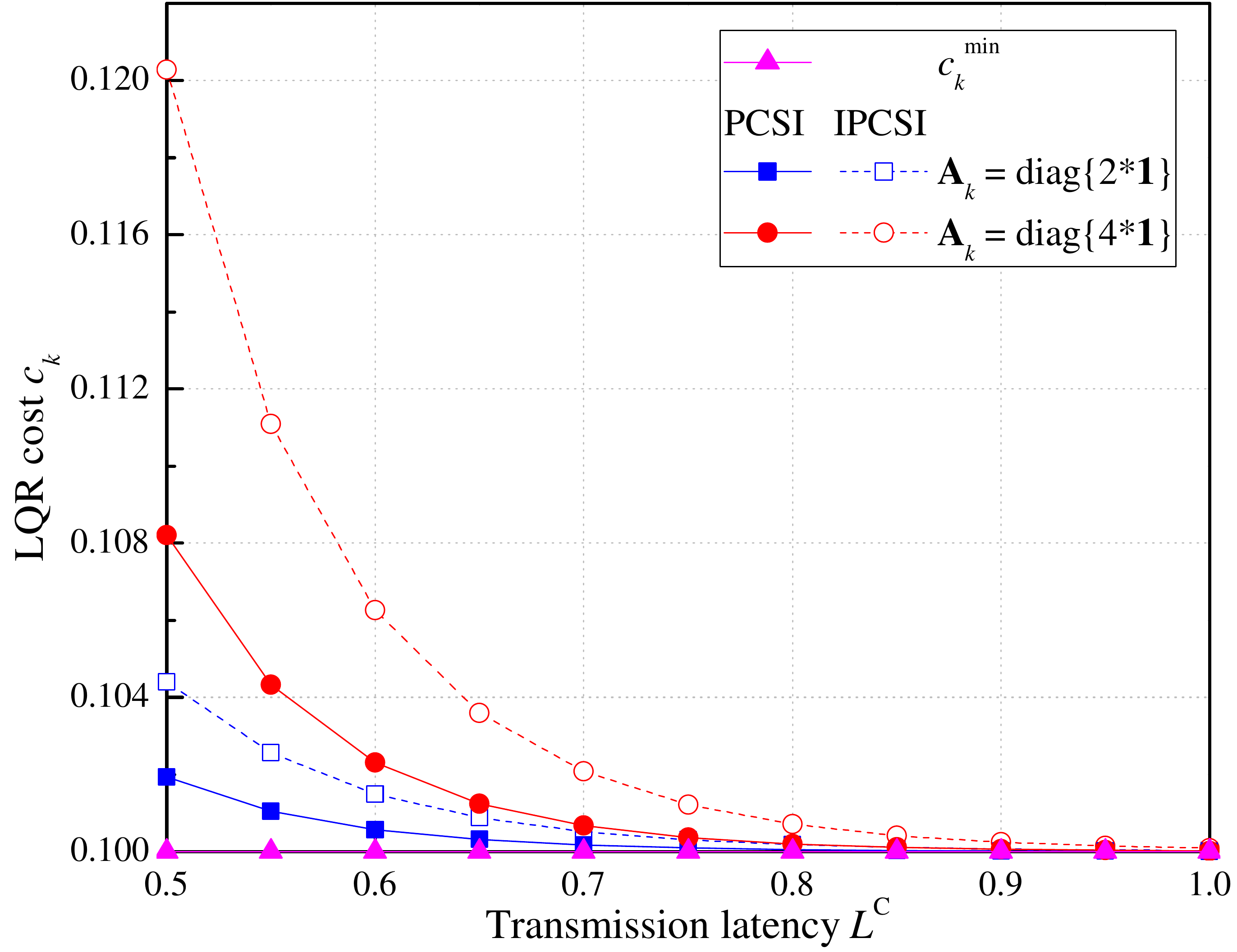}
	\caption{LQR cost versus transmission latency, with the fixed transmission power $ p^\text{C}_k = 0.5 $~\myunit{W}.}
	\label{fig_LQRvsL}
\end{figure}

\begin{figure}[!t]
	\centering
	\includegraphics[scale=0.3]{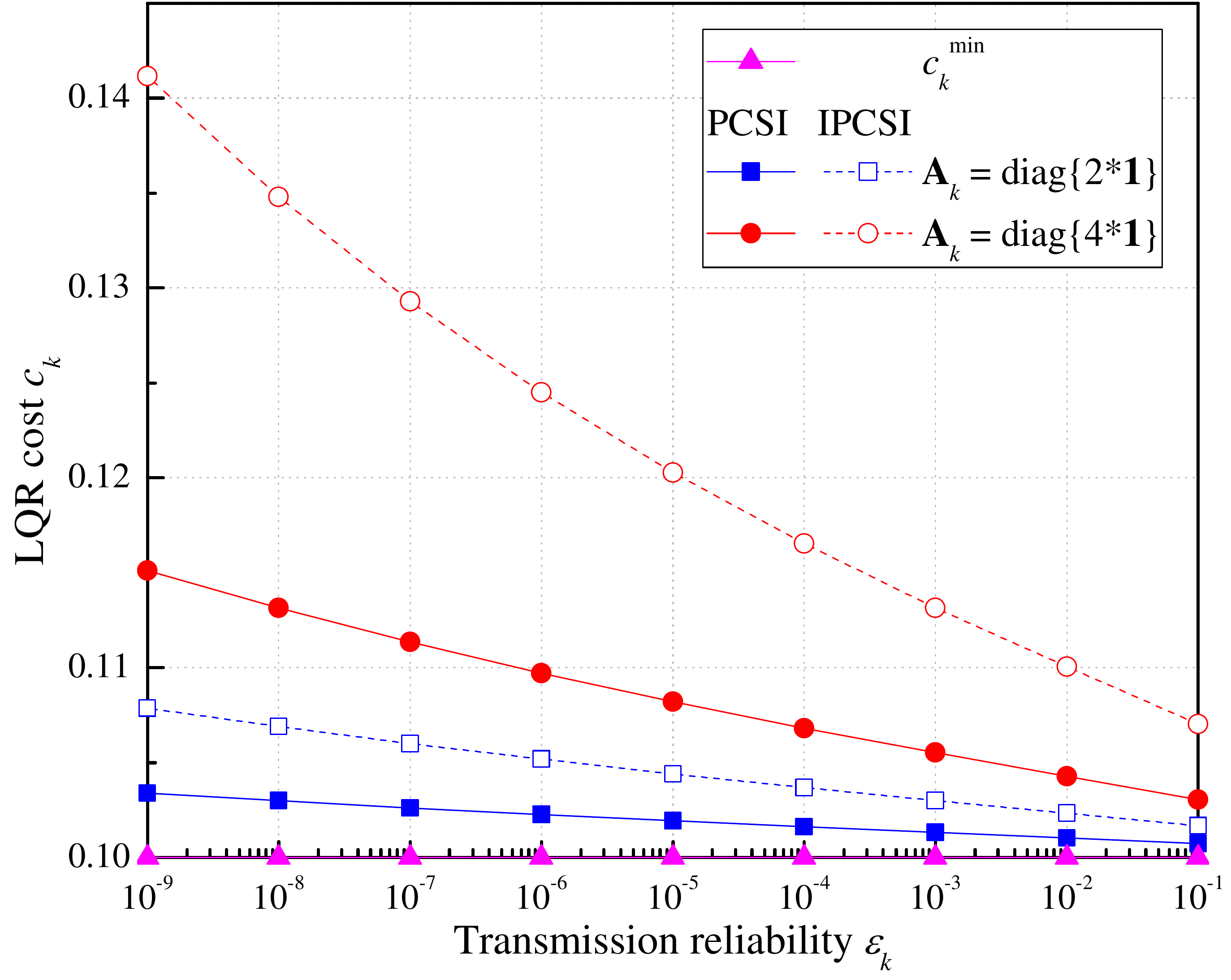}
	\caption{LQR cost versus transmission reliability, with the fixed transmission power $ p^\text{C}_k = 0.5 $~\myunit{W}.}
	\label{fig_LQRvsepsilon}
\end{figure}

\ifCLASSOPTIONtwocolumn
\begin{figure*}[!t]
	\centering
	\subfloat[Convergence of Algorithm~\ref{alg1}.]{\includegraphics[scale=0.3]{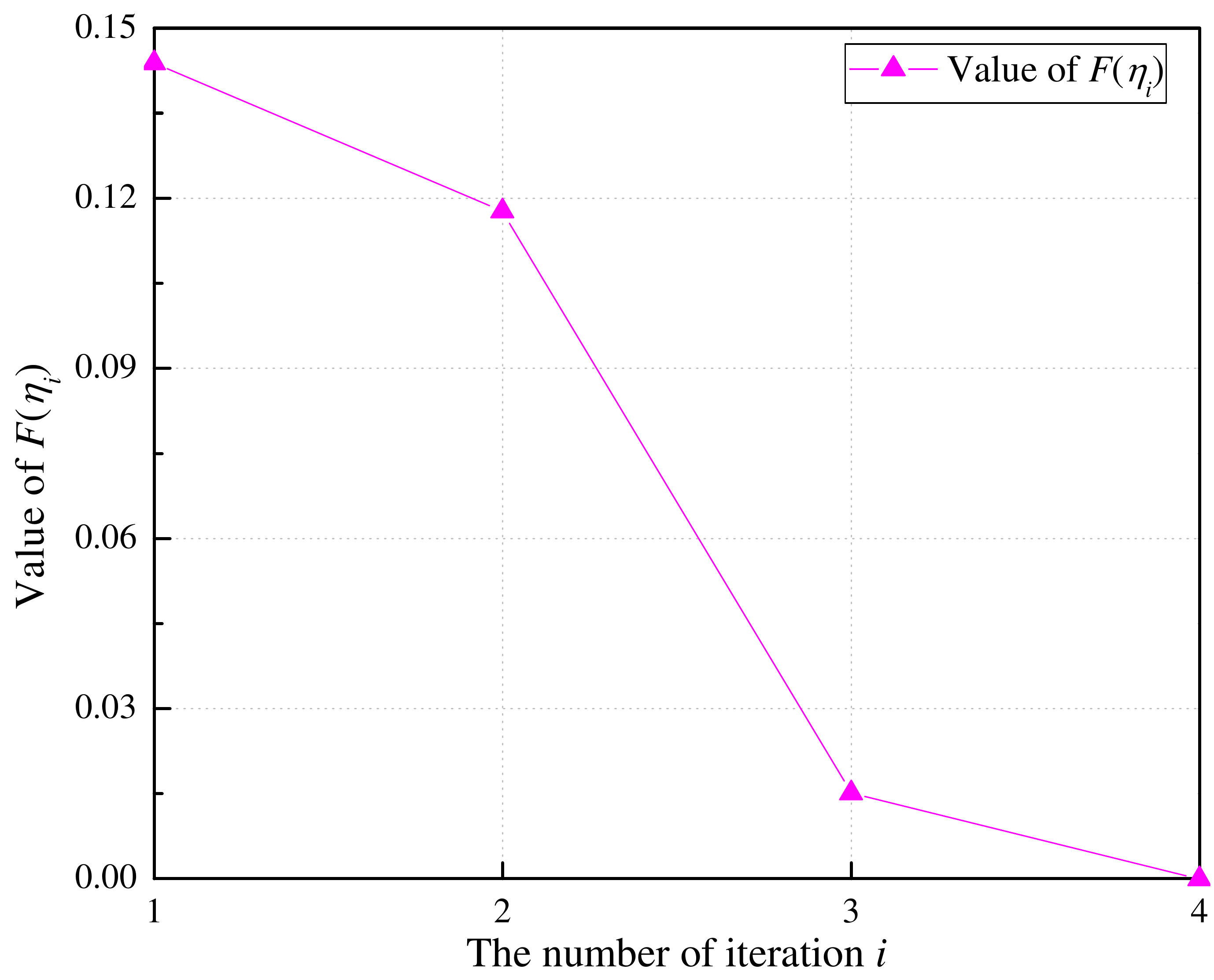}\label{fig_Cov_Alg1}}
	\hfil
	\subfloat[Convergence of Algorithm~\ref{alg2}.]{\includegraphics[scale=0.3]{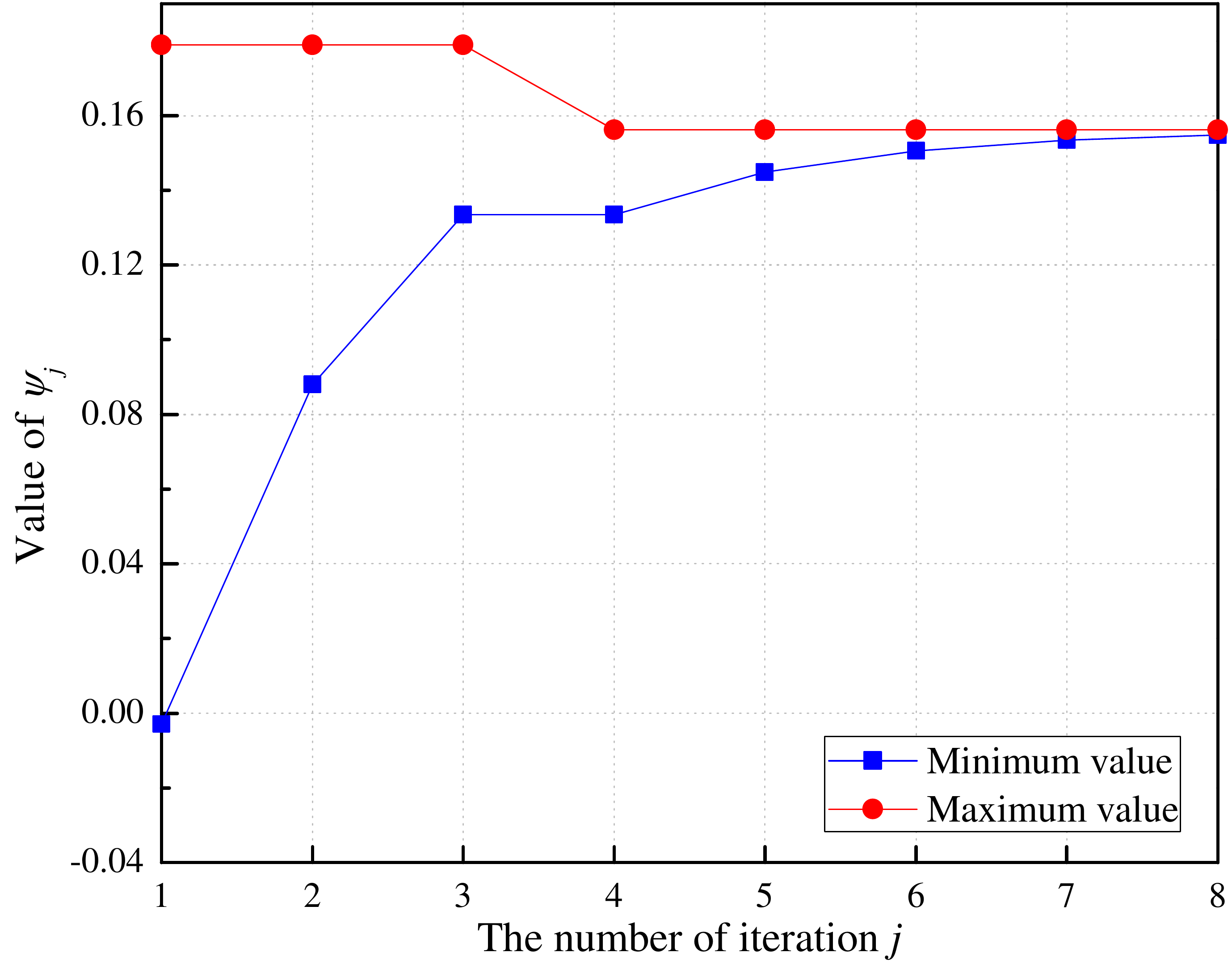}\label{fig_Cov_Alg2}}
	\caption{Convergence of proposed algorithms under IPCSI, with $ K = 8 $, $ \zeta_1 = 10^{-2} $ and $ \zeta_2 = 10^{-3} $.}
	\label{fig_Cov}
\end{figure*}

\begin{figure}[!t]
	\centering
	\includegraphics[scale=0.3]{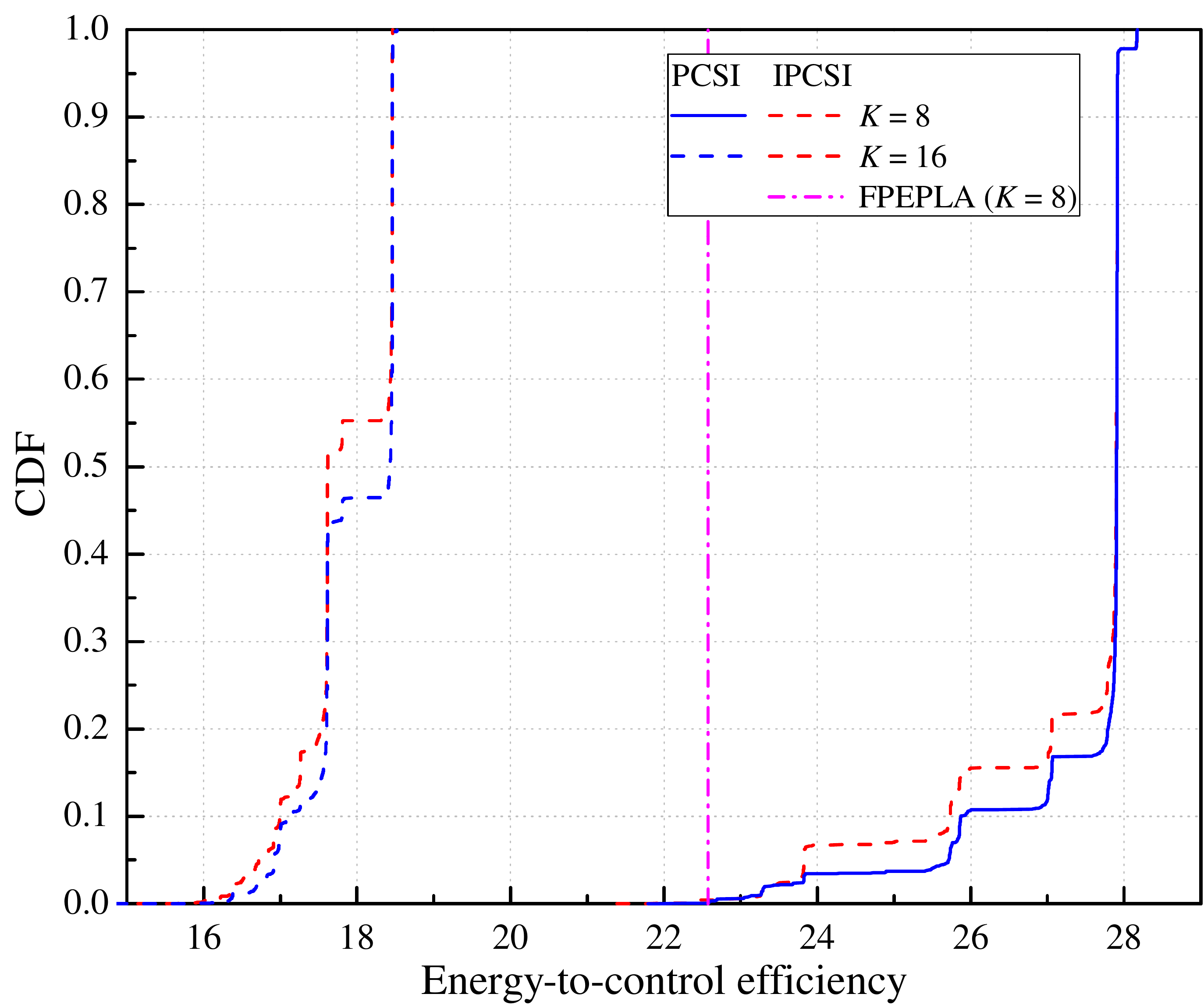}
	\caption{CDF of ECE under the different number of plants, with $ \myvec{A}_k = \diag\{2 \cdot \myvec{1}\} $, $ \rho_k = -6 $~\myunit{dB}, $ \forall k $, and the total system bandwidth $ B_\text{S} = 10 $~\myunit{MHz}.}
	\label{fig_CDFK}
\end{figure}

\begin{figure}[!t]
	\centering
	\includegraphics[scale=0.3]{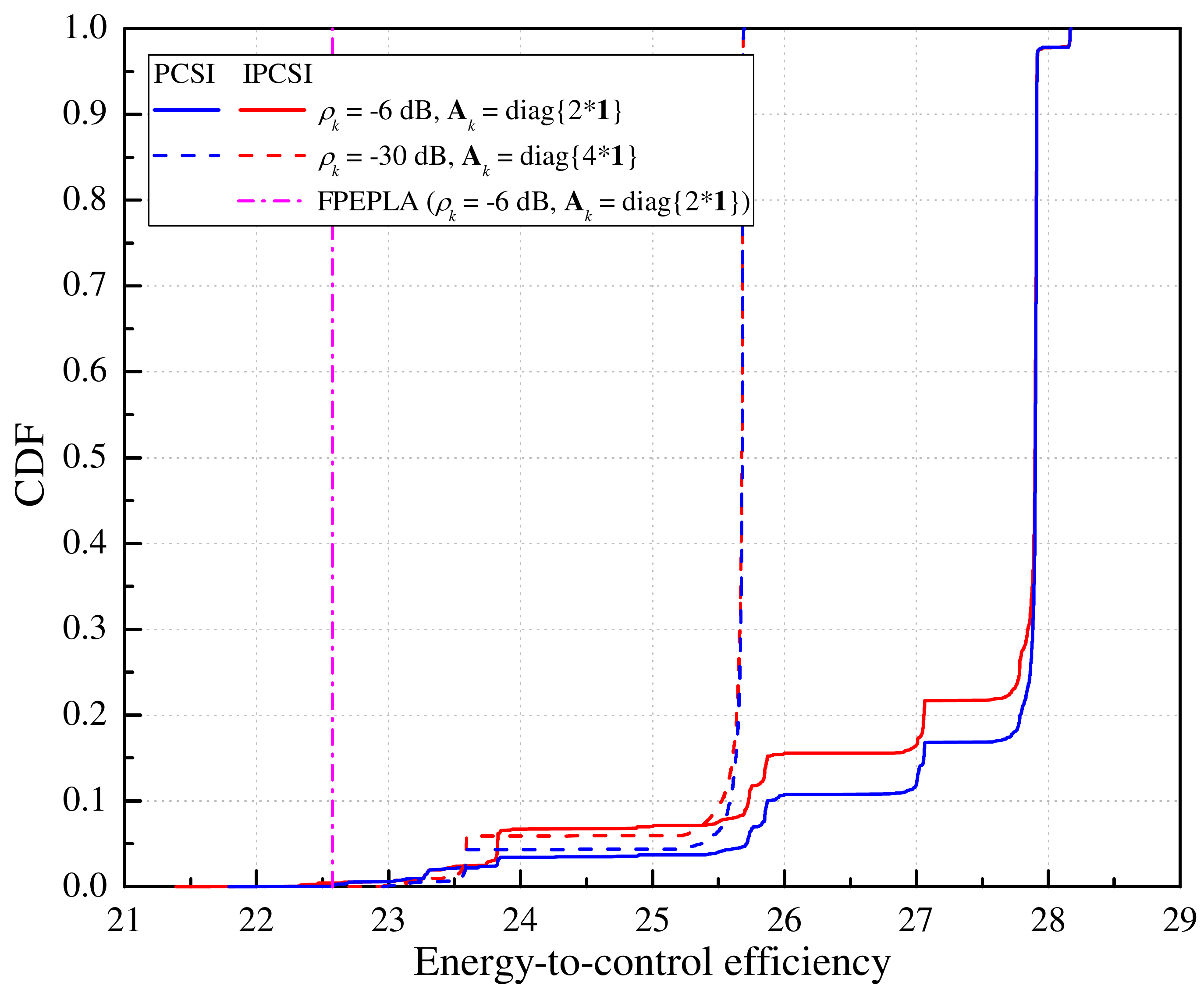}
	\caption{CDF of ECE under the different instabilities and DNRs, with $ K = 8 $ and the total system bandwidth $ B_\text{S} = 10 $~\myunit{MHz}.}
	\label{fig_CDFDNR}
\end{figure}
\fi

From Theorem~\ref{theo2}, Fig.~\ref{fig_LQRvsL} and Fig.~\ref{fig_LQRvsepsilon} illustrate the LQR cost versus the transmission latency and reliability, respectively. As shown in Fig.~\ref{fig_LQRvsL}, as the transmission latency of the control phase increases, the LQR cost decreases and finally tends to $ c^\text{min}_k $. On the contrary, Fig.~\ref{fig_LQRvsepsilon} depicts that as the transmission reliability $ \epsilon_k $ increases, the LQR cost increases. This is because increasing the transmission latency and reducing the transmission reliability can improve the amount of information during the control phase. In conclusion, Fig.~\ref{fig_LQRvsL} and Fig.~\ref{fig_LQRvsepsilon} suggest that the better LQR cost is achieved, the more amount of information is required. Furthermore, compared with the low-unstable plants ($ \myvec{A}_k = \diag\{2 \cdot \myvec{1}\} \in \mathbb{R}^{N_k \times N_k} $), the more LQR cost is paid for stabilizing the high-unstable plants ($ \myvec{A}_k = \diag\{4 \cdot \myvec{1}\} $). Finally, regardless of Fig.~\ref{fig_LQRvsL} or Fig.~\ref{fig_LQRvsepsilon}, one can indicate the imperfect CSI increases the LQR cost.

\subsection{Simulation Results for Problem~\ref{pro1}}

\subsubsection{Convergence of Proposed Algorithms}

\ifCLASSOPTIONonecolumn
\begin{figure*}[!t]
	\centering
	\subfloat[Convergence of Algorithm~\ref{alg1}.]{\includegraphics[scale=0.3]{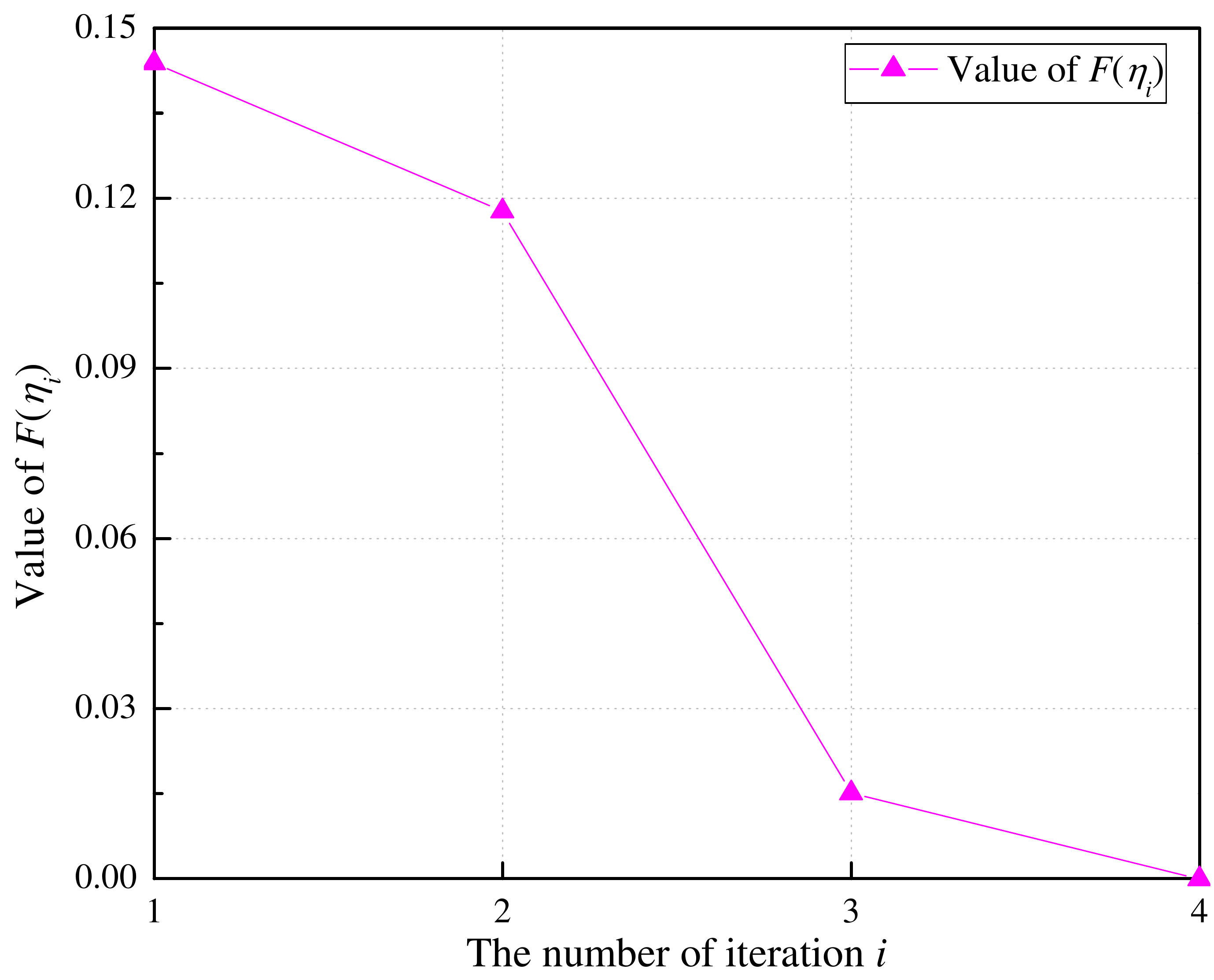}\label{fig_Cov_Alg1}}
	\hfil
	\subfloat[Convergence of Algorithm~\ref{alg2}.]{\includegraphics[scale=0.3]{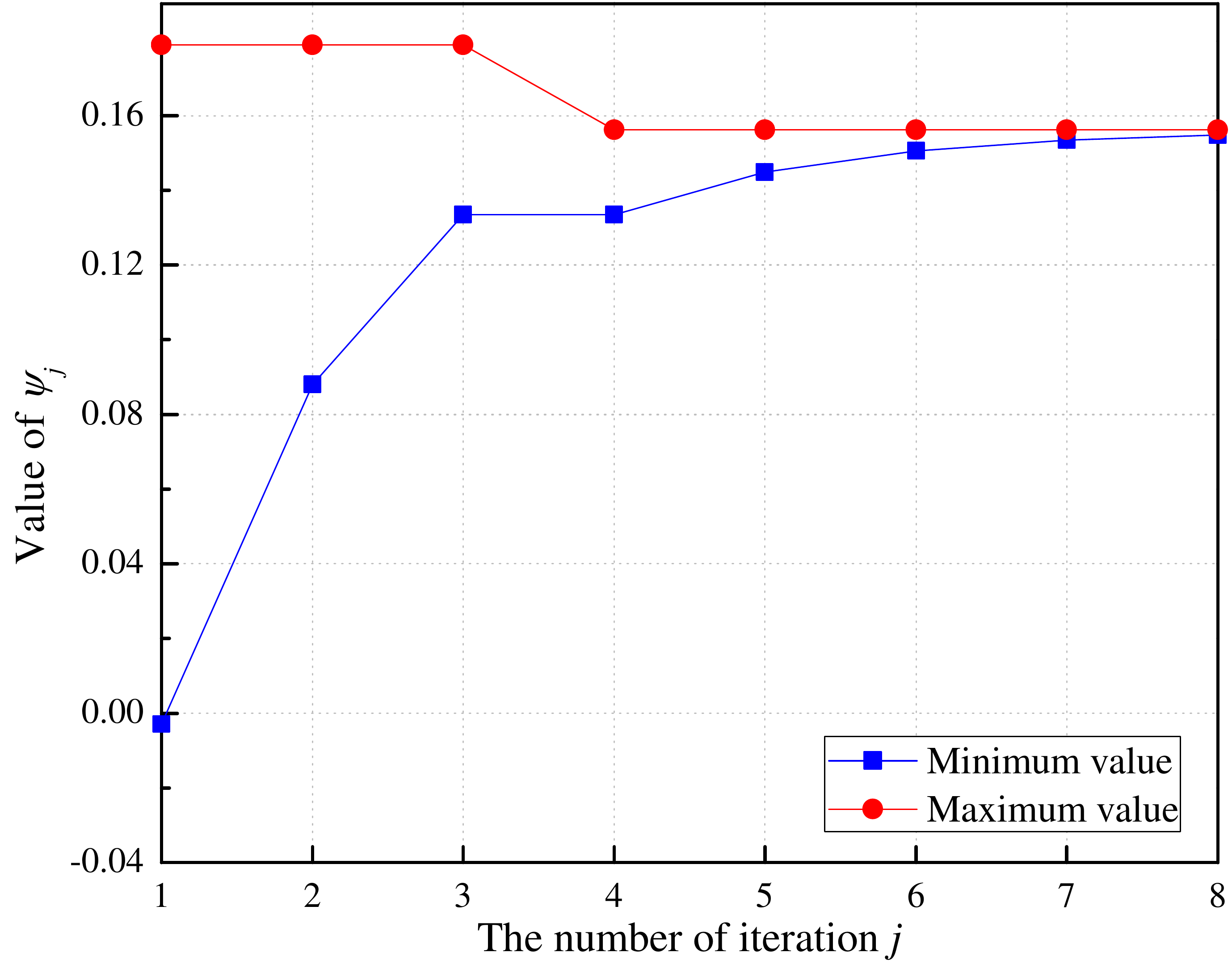}\label{fig_Cov_Alg2}}
	\caption{Convergence of proposed algorithms under IPCSI, with $ K = 8 $, $ \zeta_1 = 10^{-2} $ and $ \zeta_2 = 10^{-3} $.}
	\label{fig_Cov}
\end{figure*}
\fi

Based on a certain channel realization in the simulations, Fig.~\ref{fig_Cov} illustrates the convergence of two proposed algorithms. As shown in Fig.~\subref*{fig_Cov_Alg1}, as the number of iterations increases, the value of auxiliary function $ F(\eta_{i}) $ decreases and finally converges to zero, which is consistent with Theorem~\ref{theo5}. Thus, the convergence of Algorithm~\ref{alg1} is verified. On the other hand, Fig.~\subref*{fig_Cov_Alg2} depicts that as the number of iterations increases, the maximum value of $ \psi $ decreases and the minimum value of $ \psi $ increases. $ \psi^\text{max} $ and $ \psi^\text{min} $ of Algorithm~\ref{alg2} eventually converge to a constant. Finally, the number of iterations also indicates that Algorithm~\ref{alg1} and Algorithm~\ref{alg2} operate with the reasonable complexity.

\subsubsection{Performance of Proposed Algorithms}

\ifCLASSOPTIONonecolumn
\begin{figure}[!t]
	\centering
	\includegraphics[scale=0.3]{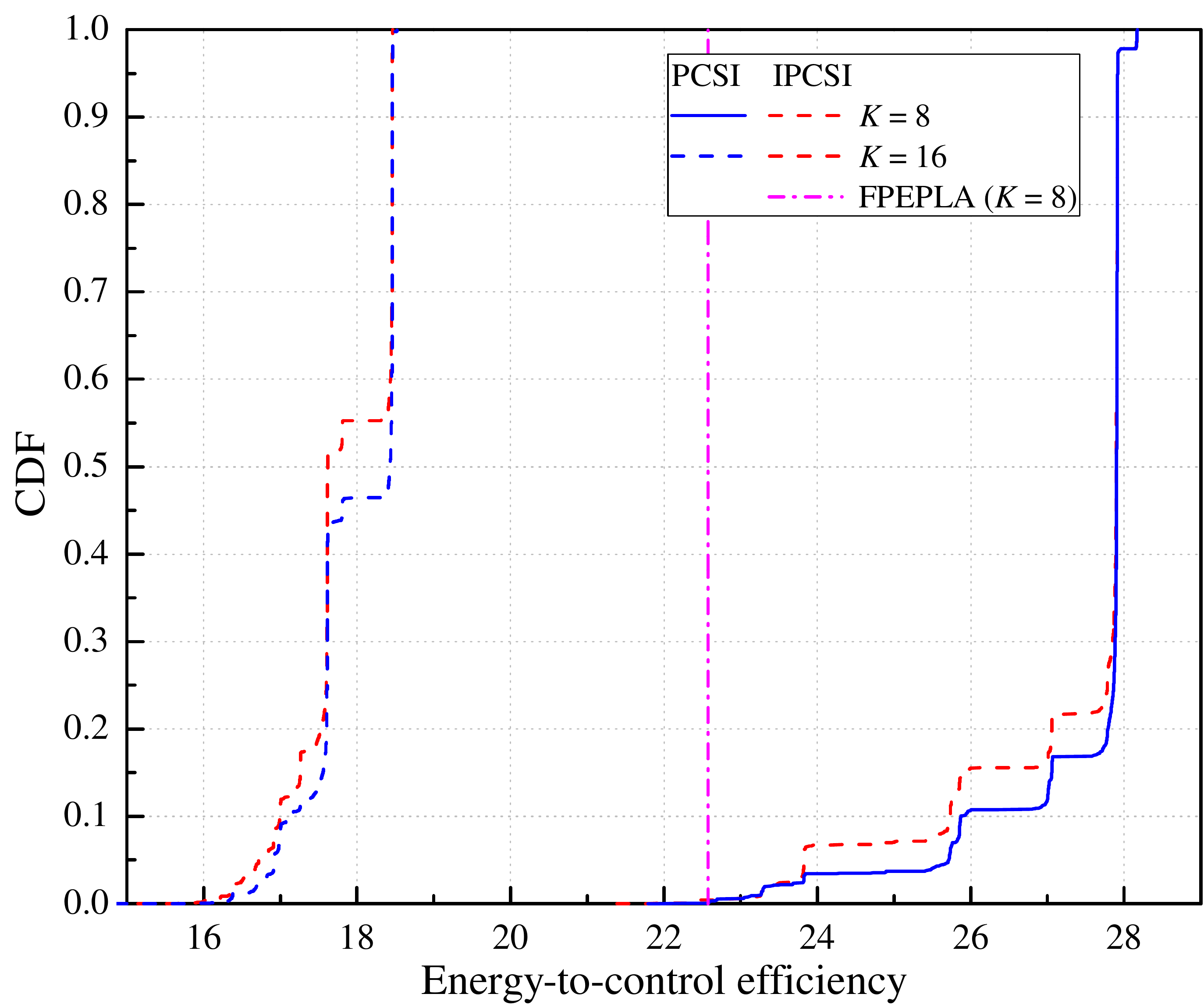}
	\caption{CDF of ECE under the different number of plants, with $ \myvec{A}_k = \diag\{2 \cdot \myvec{1}\} $, $ \rho_k = -6 $~\myunit{dB}, $ \forall k $, and the total system bandwidth $ B_\text{S} = 10 $~\myunit{MHz}.}
	\label{fig_CDFK}
\end{figure}

\begin{figure}[!t]
	\centering
	\includegraphics[scale=0.3]{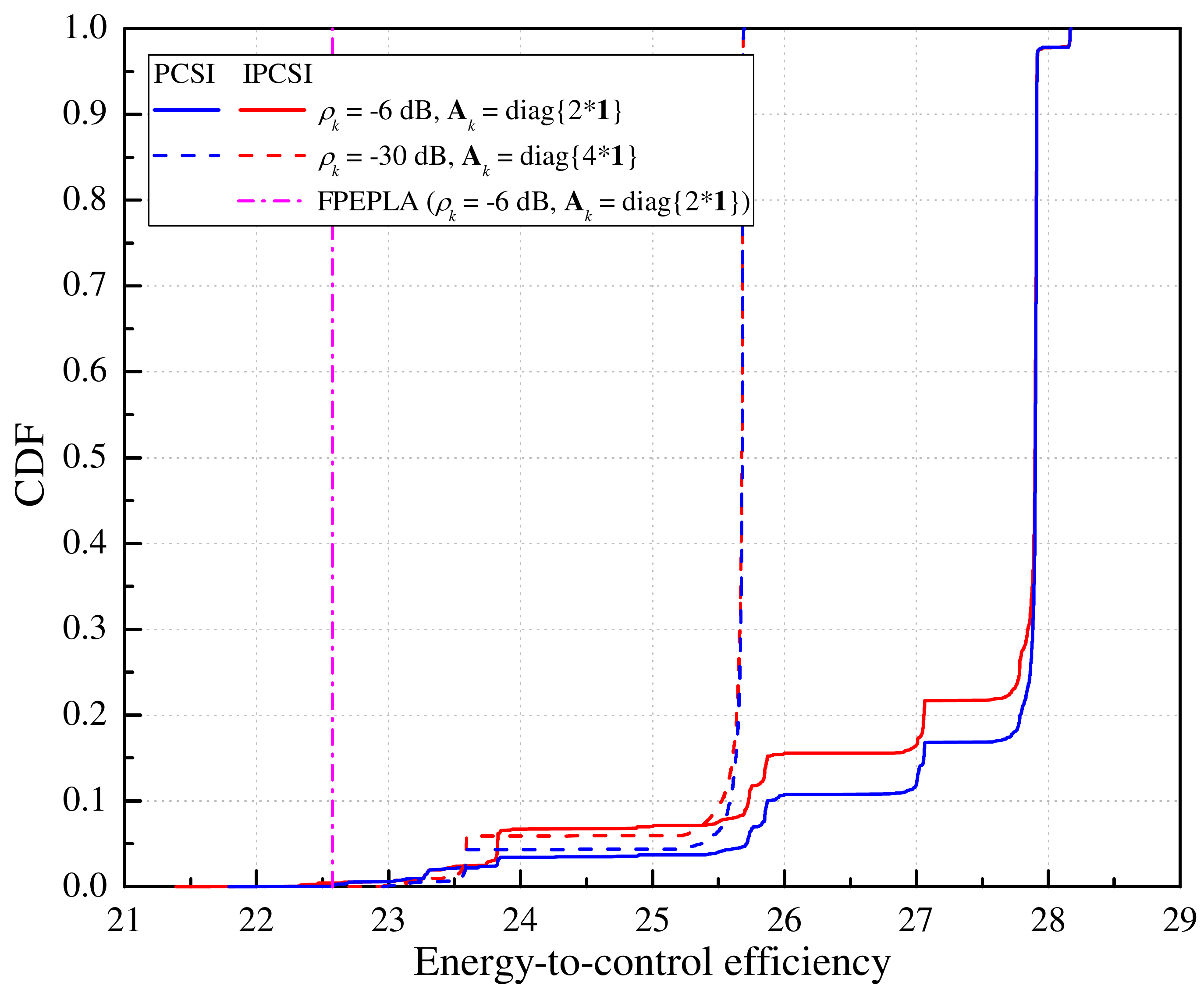}
	\caption{CDF of ECE under the different instabilities and DNRs, with $ K = 8 $ and the total system bandwidth $ B_\text{S} = 10 $~\myunit{MHz}.}
	\label{fig_CDFDNR}
\end{figure}
\fi

Let $ \rho_k = d_k/\sigma^2_\text{PS} $ denote the distortion-to-noise ratio (DNR) of the sensing phase. Fig.~\ref{fig_CDFK} illustrates the cumulative distribution function (CDF) of the ECE in Problem~\ref{pro1} under the different number of plants. As shown in Fig.~\ref{fig_CDFK}, upon increasing the number of plants (from $ K=8 $ to $ K=16 $), the performance of ECE gradually deteriorates, which means that the more plants are supported, the lower ECE is achieved. Moreover, in order to verify the performance of our proposed algorithms, the reference ECE is considered as the case with the \underline{f}ull \underline{p}ower allocation during the sensing phase and the \underline{e}qual \underline{p}ower and \underline{l}atency \underline{a}llocation during the control phase, which can be abbreviated as ``FPEPLA''. Fig.~\ref{fig_CDFK} indicates that given the number of plants ($ K=8 $), the ECE of the proposed algorithms outperforms that of FPEPLA.

Fig.~\ref{fig_CDFDNR} illustrates the CDF of the ECE in Problem~\ref{pro1} under the different instabilities and DNRs. Compared with the FPEPLA, regardless of what  the instability, DNR and CSI are, the proposed algorithms always achieve the better performance on ECE. In addition, Fig.~\ref{fig_CDFDNR} shows that for the high-unstable plants ($ \myvec{A}_k = \diag\{4 \cdot \myvec{1}\} $), the increased perception accuracy during the sensing phase (namely $ \rho_k = -30 $~\myunit{dB}) is accompanied by the decline on the performance of ECE. Finally, based on Fig.~\ref{fig_CDFK} and \ref{fig_CDFDNR}, one can demonstrate that the imperfect CSI has a very small impact on the performance of ECE. This is because when most plants operate with the high SINR, the large-scale fading is not sensitive to the imperfect CSI. To this end, we raise the following conclusions: \textit{1)} appropriately increasing the SINR during two phases is beneficial to improve the ECE; \textit{2)} our proposed algorithms can significantly improve the ECE, while guaranteeing the requirements of URLLCs for the networked control system.

\section{Conclusions}
\label{sec:Conclusion}

Networked control systems are the current trend for the emerging real-time control applications of IoT, such as autonomous driving, ``Industry 4.0'' and tactile Internet. In this paper, we investigated the optimization of the LQR cost and energy consumption for a centralized wireless networked control system operating with URLLCs. Particularly, we first developed a optimization framework including the SE during two phases, LQR cost, and energy consumption. A novel performance metric called the ECE was also proposed, and the rationality and validity of the proposed performance metric were proved. Then, with the aid of the proposed framework, we formulated a max-min joint optimization problem, and put forward two radio resource allocation algorithms to optimally solve the formulated problem. Simulation results illustrated that the proposed algorithms can greatly improve the ECE performance, while guaranteeing the requirements of URLLCs for the networked control system.

\appendices

\section{Proof of Theorem~\ref{theo1}}
\label{app:theo1}

First of all, the ergodic SE during the sensing phase can be written as
\begin{align}
\label{ergodicS}
C^\text{S2D}_k\left( \gamma_k^\text{S} \right) = \mathbb{E}_{\{\myvec{h}\}}\left[ \log_2\left( 1+\gamma_k^\text{S} \right) \right].
\end{align}
Then, according to the channel hardening, i.e., $ \hat{\myvec{g}}_{k}^\text{H} \hat{\myvec{g}}_k, \hat{\myvec{g}}_k^\text{T}\hat{\myvec{g}}_k^{*}/M \rightarrow \beta_k, \forall k $, and the asymptotic channel orthogonality, i.e., $ \hat{\myvec{g}}_{k}^\text{H} \hat{\myvec{g}}_i, \hat{\myvec{g}}_k^\text{T}\hat{\myvec{g}}_i^{*}/M \rightarrow 0, \forall i \neq k $, thus for the large number of antennas $ M $, the SINR of the $ k $-th plant during two phase can be approximated by
\begin{align}
\label{gammaS}
\gamma_k^\text{S} &= \dfrac{p^\text{S}_{k}\chi_k\beta_k}{\sigma^2_\text{S}}, M \gg 1, \\
\label{gammaC}
\gamma_k^\text{C} &= \dfrac{p_k^\text{C}\chi_k\beta_k}{\sigma^2_\text{C}}, M \gg 1.
\end{align}
Substituting~\eqref{gammaS} into \eqref{ergodicS}, we have~\eqref{SEC}.

On the other hand, in order to meet the requirements of URLLCs, the tradeoff between latency and reliability should be modeled in our optimization framework. In fact,~\eqref{setradoff} has characterized the transmission latency and the sensing-transmission reliability for the sensing phase. Hence, here we model the tradeoff between latency and reliability for the control phase. With respect to URLLCs, both the ergodic capacity and the outage capacity are no longer applicable, because they violate the requirements of URLLCs, namely $ C^\text{C2A}(\gamma^\text{C}) = \mathbb{E}_{\{\myvec{h}\}}\left[ \log_2(1+\gamma^\text{C}) \right] = \lim_{\epsilon \rightarrow 0} C^\text{C2A}_{\epsilon}(\gamma^\text{C},\epsilon) = \lim_{\epsilon \rightarrow 0}\lim_{n \rightarrow \infty} R^\text{C2A}(\gamma^\text{C},n,\epsilon) $~\cite{YangMag2019}. A common expression of $ R^\text{C2A}(\gamma^\text{C},n,\epsilon) $ is approximated by~\cite{Polyanskiy2010,She2017,Hayashi2009}:
\begin{align}
\label{ergodicC}
R^\text{C2A}\left( \gamma^\text{C},L^\text{C},\epsilon \right) \approx \mathbb{E}_{\{\myvec{h}\}} \left[ \log_2\left( 1+\gamma^\text{C} \right)-\sqrt{\dfrac{V^\text{C}}{L^\text{C}B}}Q^{-1}\left( \epsilon \right) \right],
\end{align}
where $ \epsilon $ is the proxy for the transmission reliability, and the meaning of $ V^\text{C} $ is similar to~\eqref{setradoff}. For a complex channel, the channel dispersion is given by~\cite{Hu2016}:
\begin{align}
V^\text{C} = \left( 1- \dfrac{1}{\left( 1+\gamma^\text{C} \right)^2} \right) \left( \log_2 \myexp \right)^2.
\end{align}
In the high SINR region (greater than 10~\myunit{dB}), the channel dispersion can be approximated by $ V^\text{C} \approx (\log_2 \myexp)^2 $, while in the low SINR region we have $ 0< V^\text{C} < (\log_2 \myexp)^2 $~\cite{Sebastian2015}. Therefore, $ V^\text{C} $ can be approximated by
\begin{align}
\label{approxV}
V^\text{C} \approx (\log_2 \myexp)^2,
\end{align}
which is used to obtain the lower bound in this paper. Similarly with the sensing phase, substituting~\eqref{gammaC} and \eqref{approxV} into \eqref{ergodicC}, we obtain the ergodic SE during the control phase which is given by~\eqref{SES}.

\section{Proof of Theorem~\ref{theo2}}
\label{app:theo2}

Based on Lemma~\ref{lemma1} and the source-channel separation theorem, one can conclude during the control phase
\begin{align}
L^\text{C}BC^\text{C2A}_k > R^\text{C}_k\left( c_k \right).
\end{align}
On the other hand, considering the requirements of URLLCs and $ C^\text{C2A}_k = \lim_{\epsilon \rightarrow 0} C^\text{C2A}_{\epsilon,k} = \lim_{\epsilon \rightarrow 0}\lim_{n \rightarrow \infty} R^\text{C2A}_k $, according to Theorem~\ref{theo1}, we obtain the following condition for the LQR cost, i.e.,
\begin{align}
L^\text{C}BC^\text{C2A}_k > L^\text{C}BR^\text{C2A}_k \overset{\text{(a)}}{\geqslant} R^\text{C}_k\left( c_k \right).
\end{align}
Let the equality of (a) hold and substitute~\eqref{SES} into the equality, then we have
\begin{align}
& L^\text{C}B\log_2\left( 1+\dfrac{p_k^\text{C}\chi_k\beta_k}{\sigma^2_\text{C}} \right) - \sqrt{L^\text{C}B}Q^{-1}\left( \epsilon_k \right) \log_2 \myexp \notag \\
&\quad = \log_2 \left| \det \left( \myvec{A}_k \right) \right| + \dfrac{N_k}{2}\log_2 \left( 1+ \dfrac{Z_k\left( \myvec{w}_k \right) \left| \det \left( \myvec{M}_k \right) \right|^\frac{1}{N_k}}{\frac{1}{N_k}\left( c_k-c^\text{min}_k \right)} \right),
\end{align}
which leads to~\eqref{LQRcost}.

\section{Proof of Theorem~\ref{theo3}}
\label{app:theo3}

Since the proof of~\eqref{FECE} is similar to that of~\eqref{GECE}, and the form of~\eqref{GECE} is more complex than that of~\eqref{FECE}, only the case of~\eqref{GECE} is proved here. Let
\begin{align}
\eta^\text{GECE} &= \dfrac{E^\text{GECE}_\text{max}-E^\text{S}\left( \myvec{p}^\text{S},L^\text{S} \right)-E^\text{C}\left( \myvec{p}^\text{C},L^\text{C} \right)}{\sum^{K}_{k=1} c_k\left( p_k^\text{C},L^\text{C},\epsilon_k \right)} \notag \\
&\triangleq \dfrac{f\left( \myvec{p}^\text{S},\myvec{p}^\text{C},L^\text{S},L^\text{C} \right)}{g\left( \myvec{p}^\text{C},L^\text{C} \right)}.
\end{align}
Obviously, we have
\begin{align}
\label{propf}
\begin{cases}
\dfrac{\partial^2 f\left( \myvec{p}^\text{S},\myvec{p}^\text{C},L^\text{S},L^\text{C} \right)}{\partial p_k^\text{S} \partial p_k^\text{C}} = \dfrac{\partial^2 f\left( \myvec{p}^\text{S},\myvec{p}^\text{C},L^\text{S},L^\text{C} \right)}{\partial p_k^\text{C} \partial p_k^\text{S}} = 0, \\
\dfrac{\partial^2 f\left( \myvec{p}^\text{S},\myvec{p}^\text{C},L^\text{S},L^\text{C} \right)}{\partial \left( p_k^\text{S} \right)^2} = \dfrac{\partial^2 f\left( \myvec{p}^\text{S},\myvec{p}^\text{C},L^\text{S},L^\text{C} \right)}{\partial \left( p_k^\text{C} \right)^2} = 0, \\
\dfrac{\partial^2 f\left( \myvec{p}^\text{S},\myvec{p}^\text{C},L^\text{S},L^\text{C} \right)}{\partial L^\text{S} \partial L^\text{C}} = \dfrac{\partial^2 f\left( \myvec{p}^\text{S},\myvec{p}^\text{C},L^\text{S},L^\text{C} \right)}{\partial L^\text{C} \partial L^\text{S}} = 0, \\
\dfrac{\partial^2 f\left( \myvec{p}^\text{S},\myvec{p}^\text{C},L^\text{S},L^\text{C} \right)}{\partial \left( L^\text{S} \right)^2} = \dfrac{\partial^2 f\left( \myvec{p}^\text{S},\myvec{p}^\text{C},L^\text{S},L^\text{C} \right)}{\partial \left( L^\text{C} \right)^2} = 0.
\end{cases}
\end{align}
Thus, $ f(\myvec{p}^\text{S},\myvec{p}^\text{C},L^\text{S},L^\text{C}) $ is an affine function of $ \myvec{p} = [\myvec{p}^\text{S},\myvec{p}^\text{C}] $ or $ \myvec{\tau} = [L^\text{S},L^\text{C}] $. On the other hand, because the condition $ \exp\left[ (2 \Omega_k \ln 2) / N_k \right] > 1 $ always holds in practice, we obtain
% The second order of L^\text{C}, See iPad.
\begin{align}
\label{propg}
\begin{cases}
\dfrac{\partial^2 g\left( \myvec{p}^\text{C},L^\text{C} \right)}{\partial \left( p_k^\text{C} \right)^2} > 0, \\
\dfrac{\partial^2 g\left( \myvec{p}^\text{C},L^\text{C} \right)}{\partial \left( L^\text{C} \right)^2} > 0, \\
\end{cases}
\end{align}
Hence, $ g(\myvec{p}^\text{C},L^\text{C}) $ is a convex function of $ \myvec{p}^\text{C} $ or $ L^\text{C} $. Since the properties of $ \myvec{p} = [\myvec{p}^\text{S},\myvec{p}^\text{C}] $ or $ \myvec{\tau} = [L^\text{S},L^\text{C}] $ are same for either $ f $ or $ g $, we omit all independent variables. To continue the proof, the following lemma is borrowed, i.e.,

\begin{lemma}[\cite{Boydbook,Zappone2015}]
\label{lemma4}
Let $ \mathcal{C}\subseteq\mathbb{R}^n $ be a convex set and $ r: \mathcal{C}\rightarrow\mathbb{R} $. Then, $ r $ is quasi-concave if and only if its $ \xi $-superlevel set $ \mathcal{S}_\xi=\left\lbrace \myvec{a} \in \mathcal{C} : r(\myvec{a}) \geqslant \xi \right\rbrace $ is convex for all $ \xi \in \mathbb{R} $.
\end{lemma}

Based on Lemma~\ref{lemma4}, $ \mathcal{S}_\xi = \left\lbrace f / g \geqslant \xi \right\rbrace $ is the empty set for all $ \xi \leqslant 0 $ when $ f>0 $ and $ g>0 $, thus only the case of $ \xi > 0 $ is considered. The equivalent form of $ \mathcal{S}_\xi $ is given by $ \mathcal{S}_\xi=\left\lbrace f - \xi g \geqslant 0 \right\rbrace $. Let $ u = f - \xi g $, then $ u^{\prime\prime} = f^{\prime\prime} - \xi g^{\prime\prime} < 0 $. Therefore, $ u = f - \xi g $ is concave, i.e., the $ \xi $-superlevel set $ \mathcal{S}_\xi $ is convex, and $ \eta^\text{GECE} $ is a quasi-concave function of $ \myvec{p} = [\myvec{p}^\text{S},\myvec{p}^\text{C}] $ or $ \myvec{\tau} = [L^\text{S},L^\text{C}] $.

Subsequently, we prove that $ \eta^\text{GECE} $ is also a pseudo-concave function of $ \myvec{p} = [\myvec{p}^\text{S},\myvec{p}^\text{C}] $ or $ \myvec{\tau} = [L^\text{S},L^\text{C}] $. We take the power vector $ \myvec{p} $ as an \textit{example}. Let $ \myvec{p}^* $ be a stationary point of $ \eta^\text{GECE}(\myvec{p}) = f(\myvec{p})/g(\myvec{p}) $, then we have $ \nabla f(\myvec{p}^*) = \eta^\text{GECE}(\myvec{p}^*)\nabla g(\myvec{p}^*) $. Moreover, from~\eqref{propf} and \eqref{propg} we have
\begin{align}
f\left( \myvec{p} \right) &= f\left( \myvec{p}^* \right) + \nabla f(\myvec{p}^*)^\text{T} \left( \myvec{p}-\myvec{p}^* \right) \notag \\
&= f\left( \myvec{p}^* \right) + \eta^\text{GECE}(\myvec{p}^*)\nabla g(\myvec{p}^*)^\text{T} \left( \myvec{p}-\myvec{p}^* \right) \notag \\
&< f\left( \myvec{p}^* \right) + \eta^\text{GECE}(\myvec{p}^*) \left[ g\left( \myvec{p} \right) - g\left( \myvec{p}^* \right) \right] \notag \\
&= \eta^\text{GECE}(\myvec{p}^*)g\left( \myvec{p} \right),
\end{align}
namely $ \eta^\text{GECE}(\myvec{p}) < \eta^\text{GECE}(\myvec{p}^*) $, which means that the stationary point $ \myvec{p}^* $ is the local maximum. To this end, $ \eta^\text{GECE} $ is a pseudo-concave function of $ \myvec{p} = [\myvec{p}^\text{S},\myvec{p}^\text{C}] $ or $ \myvec{\tau} = [L^\text{S},L^\text{C}] $.

Finally, $ \eta^\text{FECE} $ can only be a quasi-concave function of $ \myvec{p} = [\myvec{p}^\text{S},\myvec{p}^\text{C}] $ or $ \myvec{\tau} = [L^\text{S},L^\text{C}] $. This is because the function $ \min(\cdot) $ is not differentiable.

%% *************************************************************************
%\section*{Acknowledgment}
%\addcontentsline{toc}{section}{Acknowledgment}
%
%\blindtext
%% *************************************************************************
%% References section
%%
%% Can use a bibliography generated by BibTeX as a .bbl file.
%%
\bibliographystyle{IEEEtran}
%% Argument is your BibTeX string definitions and bibliography database(s).
\bibliography{IEEEabrv,Ref}
%%
%% <OR> Manually copy in the resultant .bbl file.
%% Set second argument of \begin to the number of references.
%% (used to reserve space for the reference number labels box)
%%
%\begin{thebibliography}{1}
%	
%	\bibitem{IEEEhowto:kopka}
%	H.~Kopka and P.~W. Daly, \emph{A Guide to {\LaTeX}}, 3rd~ed.\hskip 1em plus
%	0.5em minus 0.4em\relax Harlow, England: Addison-Wesley, 1999.
%	
%\end{thebibliography}
%%
%% *************************************************************************
%\begin{IEEEbiography}[{\includegraphics[width=1in,height=1.25in,clip,keepaspectratio]{HaojunYang}}]{Haojun Yang}
%\begin{IEEEbiography}{Haojun Yang}
%\blindtext
%\end{IEEEbiography}
%
%\begin{IEEEbiography}{Kan Zheng}
%\blindtext
%\end{IEEEbiography}
%% *************************************************************************
%% End All
\end{document}